\documentclass[
 aip,
 reprint, apl,
 superscriptaddress,
 citeautoscript,
]{revtex4-2}
\usepackage[subpreambles=false]{standalone}
\usepackage{import}
\usepackage[utf8]{inputenc}
\usepackage[T1]{fontenc}
\usepackage{cmap}
\usepackage{graphicx}
\usepackage{mathtools}

\usepackage[hyperfootnotes=false,breaklinks=true]{hyperref}
\usepackage{setspace}
\hypersetup{
 bookmarksopen=true,
 bookmarksopenlevel=1,
 colorlinks=true,
 linkcolor=cyan,
 anchorcolor=cyan,
 citecolor=cyan,
 filecolor=cyan,
 urlcolor=cyan,
 pdfpagemode=UseOutlines,
 pdfstartview={XYZ null null 1},
 linktocpage=true,
}
\usepackage{amsmath}
\allowdisplaybreaks  
\usepackage{amssymb}
\usepackage{amstext}
\usepackage{amsfonts}
\usepackage{mathrsfs}
\usepackage{amsthm}
\usepackage{dsfont}
\usepackage{bm}
\usepackage{braket}
\usepackage{physics}
\usepackage{enumitem}
\usepackage[
 capitalise,
]{cleveref}
\crefname{table}{Table}{Tables}%
\crefname{section}{Sec.}{Sects.}%
\Crefname{section}{Section}{Sections}%
\Crefname{table}{Table}{Tables}%

\usepackage{pgfplots}
\pgfplotsset{compat=newest}
\usepackage{tikz}
\usetikzlibrary{
  positioning,
  shapes.geometric,
  arrows,
  arrows.meta,
  backgrounds,
  fit,
  intersections,
  automata,
}
\usepackage{qcircuit}
\usepackage{algorithm,algorithmic}

\usepackage[draft, commentmarkup=uwave]{changes}
\definechangesauthor[name={Sarah Chehade}, color=orange]{SC}
\definechangesauthor[name={Eugene Dumitrescu}, color=red]{EFD}
\definechangesauthor[name={Yan Wang}, color=cyan]{YW}
\newcommand{\stkout}[1]{\ifmmode\text{\sout{\ensuremath{#1}}}\else\sout{#1}\fi}
\setdeletedmarkup{\stkout{#1}}

\theoremstyle{plain}

\theoremstyle{definition}

\theoremstyle{remark}

\newcommand{\mycomment}[1]{}

\newcommand{\Arg}{\operatorname{Arg}}

\newcommand{\Prob}{\operatorname{\textbf{Prob}}}

\newcommand{\mbb}[1]{\mathbb{#1}}

\newcommand{\bLozenge}{\mathbin{\blacklozenge}}
\newcommand{\tagthis}{\addtocounter{equation}{1}\tag{\theequation}}

\graphicspath{{.}{./figs/}}

\usepackage{relsize}
\usepackage{cancel}

\makeatletter
\def\blfootnote{\xdef\@thefnmark{}\@footnotetext}
\makeatother

\usepackage{subfiles}
\usepackage{orcidlink}

\usepackage{booktabs}
\usepackage{multirow, makecell}

\usepackage{pifont}
\newcommand{\cmark}{\ding{52}}%
\newcommand{\xmark}{\ding{56}}%

\usepackage[rightcaption]{sidecap}
\sidecaptionvpos{figure}{c}

\begin{document}

\blfootnote{This manuscript has been authored by UT-Battelle, LLC, under Contract No. DE-AC0500OR22725 with the U.S. Department of Energy. The United States Government retains and the publisher, by accepting the article for publication, acknowledges that the United States Government retains a non-exclusive, paid-up, irrevocable, worldwide license to publish or reproduce the published form of this manuscript, or allow others to do so, for the United States Government purposes. The Department of Energy will provide public access to these results of federally sponsored research in accordance with the DOE Public Access Plan.}

\title{Semicoherent symmetric quantum processes: Theory and applications}

\author{Yan Wang\,\orcidlink{0000-0002-6545-6434}}
\email{wangy2@ornl.gov}

\author{Sarah Chehade\,\orcidlink{0000-0003-3130-1247}}
\email{chehades@ornl.gov}

\author{Eugene Dumitrescu\,\orcidlink{0000-0001-5851-9567}}
\email{dumitrescuef@ornl.gov}

\affiliation{Computational Sciences and Engineering Division, 
Oak Ridge National Laboratory, 
Oak Ridge, Tennessee 37831, USA}

\date{\today}

\begin{abstract}
Discovering pragmatic and efficient approaches to construct $\varepsilon$-approximations of quantum operators such as real (imaginary) time-evolution propagators in terms of the basic quantum operations (gates) is challenging. Prior $\varepsilon$-approximations are invaluable, in that they enable the compilation of classical and quantum algorithm modeling of, e.g., dynamical and thermodynamic quantum properties. In parallel, symmetries are powerful tools concisely describing the fundamental laws of nature; the symmetric underpinnings of physical laws have consistently provided profound insights and substantially increased predictive power. In this work, we consider the interplay between the $\varepsilon$-approximate processes and the exact symmetries in a semicoherent context---where measurements occur at each logical clock cycle. We draw inspiration from Pascual Jordan's groundbreaking formulation of nonassociative, but commutative, symmetric algebraic form. Our symmetrized formalism is then applied in various domains such as quantum random walks, real-time evolutions, variational algorithm ansatzes, and efficient entanglement verification. Our work paves the way for a deeper understanding and greater appreciation of how symmetries can be used to control quantum dynamics in settings where coherence is a limited resource.
\end{abstract}

\maketitle


\section{Introduction}
Block encoded quantum operators enable linear combination of unitaries (LCU)~\cite{Childs2012}, as well as $\mathfrak{su}(2)$~\cite{Low2017} and $\mathfrak{su}(1,1)$~\cite{Rossi2023} quantum signal processing. They also facilitate  the generalizations of $\mathfrak{su}(2)$ qubitization~\cite{Low2019}, and more recently, simulations of open quantum dynamics~\cite{Hu2020, Suri2023}. 
Numerous optimizations of these algorithms exist~\cite{Low2019a, CarreraVazquez2023, Rendon2022, Zhuk2023}. When used in a \textit{fully coherent} manner, quantum resources offer theoretical gold-standard speedups with respect to precision (i.e., error $\varepsilon$) scalings~\cite{Martyn2023}. However, the resources necessary to dynamically evolve the system over long time scales in a fully coherent manner are not currently available in present or near-term quantum devices. This motivates us to take a more pragmatic approach and ask: What applications are realizable with more near-term \emph{semicoherent} quantum dynamics due to quantum channels with interleaved unitaries and measurement operations?

Hale Trotter's operator splitting decomposition method~\cite{Trotter1959, Suzuki1976} for approximating noncommutative semigroup dynamics is a natural starting choice for integrating the unitary Lie group dynamics of the Schr\"odinger equation $i\hbar\frac{\partial}{\partial t} \ket{\Psi(t)} = H \ket{\Psi(t)}$ and factorizing the nonunitary thermal density operator $e^{-\tau H}$ for imaginary time evolutions, where $\tau = \beta = 1/T$ the inverse temperature. In both cases, the Hamiltonian operator $H$ can be split into at least two terms $H = H_A + H_B$, where the \textit{individual} factors $H_A$ and $H_B$ can be exactly or efficiently exponentiated. Despite of such a simple recipe from decades ago, amidst the second quantum revolution the study of Trotter factorization errors~\cite{Childs2021} has continued to exponentially grow like Hilbert spaces. 
Hence, it is desirable to develop an analytical theory for Trotter formulas and optimize them with regard to precision and cost. This applies to applications in both classical and quantum computing~\cite{Haah2021, Ostmeyer2023, Zeng2022, Zhuk2023}.
Throughout these studies, Trotter formulas often exhibit a time-reversal inversion symmetric partner $U(-t)$ such that $U(-t) U(t) = \mathds{1}$, which corresponds to substituting $-t$ for $t$ in the approximate integrator\footnote{We use the term ``integrator'' for the propagator for both classical Hamiltonians~\cite{Yoshida1990} and quantum Hamiltonians. In the latter case, it becomes the time-evolution operator.} $U(t)$, formulated by Suzuki~\cite{Suzuki1991} (see Theorem 2 therein). 

\begin{figure}
 \centering
 \includegraphics[width=\columnwidth]{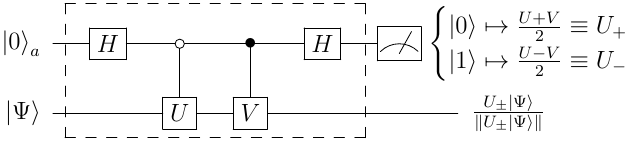}
\caption{Selecting $U=V^\dagger=e^{-itH}$ results in a spectral random walk (\cref{sec:apps-Qwalk}). Alternatively, setting $U= e^{tA} e^{tB}$ and $V = e^{tB} e^{tA}$, the quantum circuit acts by the BCH-like series that is symmetric with respect to the inversion $A\leftrightarrow B$ and Trotter-factorizes time-evolution (\cref{sec:apps-TrotterFactorization}). Setting different times ($t\rightarrow \theta_1, \theta_2$) enables a symmetric variational ansatze (\cref{sec:apps-symHVA}). Last, setting $U=X_l$ and $V=iY_l$ for an array of qubits $\{q_l\}_{l=1}^N$ and concatenating the gadget $N$-times performs the measurement of Mermin polynomial $M_N$ with a linear depth circuit (\cref{sec:apps-MerminPoly}). The symmetry of the operator applied to $\ket{\Psi}$ is contingent on the ancillary qubit's measurement outcome, that is, the $\ket{1}$ or $\ket{0}$ state. Note that $U_\pm$ are not unitary and that the principal system's final state $U_\pm \ket{\Psi}$ is normalized upon measurement of the ancilla qubit, due to the measurement postulate.}
 \label{fig:circuit}
\end{figure}

In this work, we apply symmetric forms\footnote{This includes both symmetry and anti-symmetry, associated with $+1$ and $-1$ eigenvalue, respectively.} that combine important features of block encodings and algebraic symmetries to quantum processes spanning from dynamics to entanglement verification and obtain compilations that are suitable for near-term quantum simulations. This work is organized as follows.
\Cref{sec:factorization} defines the algebraic methodologies, culminating \cref{eq:z+_4th_order}, which is the generator of the \emph{symmetry gadget} depicted by the quantum circuit in \cref{fig:circuit}. The symmetry gadget can be concatenated in iterative algorithms even though the Jordan product $\circ$, on which the gadget is based, is nonassociative.
In \cref{sec:apps}, we adapt our protocol for a variety of quantum computing applications through variations of the symmetry gadget. First, we synthesize symmetrized nonunitary time-evolution operators, such as $\frac{1}{2}(e^{-itH}\pm e^{+itH})$ [i.e., ${\cos}(tH)$ or ${\sin}(tH)$], to naturally encode semicoherent symmetric quantum walks (\cref{sec:apps-Qwalk}). Through projective measurements, the gadget steers the quantum state to eigenstates of the target Hamiltonian since they are the absorbing states of the symmetrical-walk and thus can be used to prepare eigenstates. Next, we propose a second-order Trotter form given by the \emph{commutative} Jordan product~\cite{Jordan1932}: $e^{itH_A} \circ e^{itH_B} \equiv  \frac{1}{2}(e^{itH_A}e^{itH_B} + e^{itH_B}e^{itH_A})$ (\cref{sec:apps-TrotterFactorization}). By preserving the algebraic inversion symmetry with respect to exchanging the $A$ and $B$ subparts (sublattices) we show systematic improvements in factorizations of random and deterministic dynamics and also in variational ansatz convergence (\cref{sec:apps-symHVA}).
Reducing the measurement complexity of entanglement verification protocols is the last application presented in \cref{sec:apps-MerminPoly} before a concluding discussion in \cref{sec:conclusion}. 

\section{Exchange Symmetric Factorization} \label{sec:factorization}

\subsection{Higher-Order Precision Extrapolation} \label{sec:generator}
To simplify notation we write the generator $H = A + B$ in a two-term split form. To approximate the global integrator as a function of the time $t$ parameter, with split-term integrators $e^{tA}$ and $e^{tB}$, the first order Trotter formula reads $e^{tH} \approx S_1(t) + \order*{t^2[A,B]}$, where $S_1(t) = e^{tA} e^{tB}$ or $e^{tB} e^{tA}$. To synthesize more accurate approximations, one idea is to combine counter-driving time-evolution terms that destructively interfere with the error contributions at leading orders in the operator expansion.
%
These additional time evolution terms can be introduced at the end of the complete time evolution path or sandwiched between fractional points along the path. Moreover, they can be applied in either additive or multiplicative forms to eliminate leading-order commutators.
For example, the well-known Zassenhaus formula~\cite{magnus1954exponential} adds a multiplicative correction to the end of the path. On the other hand, the Lie--Trotter--Suzuki formulas~\cite{Suzuki1991} incorporate multiplicative correction throughout a fractional path, e.g., $S_2(t) = e^{(t/2)A} e^{tB} e^{(t/2)A}$, which restores the time-reversal symmetry (TRS) $S_2(-t) S_2(t) = \mathds{1}$ that is only preserved up to $\order{t}$ in $S_1(t)$. However, $S_2$ is still not invariant under $A \leftrightarrow B$, lacking the exchange symmetry with respect to subparts.

Therefore, to increase the order of the Trotter formula and simultaneously reduce fluctuations from truncated higher orders, which contribute to the overall error norm, we investigate symmetrizing across $A$ and $B$ using either the (anti-commutator) additive form $e^{tA} e^{tB} + e^{tB} e^{tA} \equiv \{e^{tA}, e^{tB} \}$ or the commutator form $e^{tA} e^{tB} - e^{tB} e^{tA} \equiv [e^{tA}, e^{tB}]$. The proposed methods are investigated in details below.

First, recall that the Baker--Campbell--Hausdorff (BCH) formula is a formal Lie series solution to $z$ in the equation $e^{A} e^{B} = e^{z(A,B)}$, where $A$ and $B$ do not necessarily commute. Since the group-action inverse of $e^{z(A,B)}$ is $e^{-z(A,B)} = [e^{z(A,B)}]^{-1} = (e^{A} e^{B})^{-1} = e^{-B} e^{-A} = e^{z(-B,-A)}$ (last equality simply by definition), an algebraic symmetry of the BCH series expansion exists, namely, $-z(A, B) =  z(-B, -A)$. This symmetry manifests as a combination of time-reversal and inversion symmetry (TRIS), expressed by the following equation:
\begin{align}
  -z(tA, tB) &= z((-t)B, (-t)A),
\end{align}
whenever a time variable $t$ is explicitly introduced. Equivalently, the TRIS is also expressed as $z(A,B) + z(-B,-A) = 0$. Note that the BCH formula is not symmetric under either time-reversal (TR: $t\to-t$) or inversion (I: $A\leftrightarrow B$) individually but only under the product TRIS. For a bipartite lattice, the inversion (I) coincides with the spatial inversion.

The $-t$ reversed-time (i.e., backward-time) evolution often appears as intermediate time-steps in the higher-order $\varepsilon$ approximations. The Lieb--Robinson--Haah~\cite{Haah2021} construction utilizes backward-time evolution to minimize approximation errors. Backward-time evolution is also necessary in Lie--Trotter--Suzuki formulas, as highlighted by Theorem~3 in Ref.~\onlinecite{Suzuki1991} (except for the first and second order Trotter formulas). In examples involving LCU, certain coefficients within the linear combination must be negative~\cite{Childs2012} as well. This introduces instability in the success probability of the unitary approximations by LCU.

By expressing $z(A,B) = \sum_{n=1}^\infty z_n(A,B)$, where $z_n(A,B)$ is a homogeneous degree-$n$ \emph{Lie polynomial} that can be expressed in terms of $(n-1)$-level nested commutators of $A$ and $B$, we derive some key relationships resulted from TRIS. Specifically, the sum of the odd-degree terms of the expansion, $z_\text{odd}(A,B) = \sum_\text{odd $n$} z_n(A,B)$, has the I-symmetry $z_\text{odd}(A,B) = z_\text{odd}(B,A)$ due to the I-symmetry $z_n(A,B) = z_n(B,A)$ for all odd $n$. Meanwhile, the sum of the even-degree terms, $z_\text{even}(A,B)$, fulfills an inversion anti-symmetry $z_\text{even}(A,B) = -z_\text{even}(B,A)$, as a consequence of $z_n(A,B) = -z_n(B,A)$ for all even $n$. The algebraic symmetry $z_n(A,B) = (-1)^{n-1} z_n(B,A)$ for all integer $n$ is a part of Theorem~1 in Goldberg's work~\cite{Goldberg1956}. Another way to see this is to explicitly factor out the abelian time-path $-t$ factor; noting the TRIS must also hold for every order of $t^n$. That is, $z_n(B,A) t^n = z_n(tB, tA) \stackrel{\text{TRIS}}{=} -z_n ((-t)A,(-t)B) = -(-t)^n z_n(A, B) = (-1)^{n-1} z_n (A,B) t^n$, so that $z_n(A,B) = (-1)^{n-1} z_n(B,A)$.

The symmetry relations can be explicitly verified for the leading terms of the BCH series \cref{eq:BCH_series} (see, e.g., Ref.~\onlinecite{Bonfiglioli2012}),
\begin{subequations}
\label{eq:BCH_series}
\begin{align}
  &\phantom{={}}  z(A, B) 
\equiv A \bLozenge B \equiv \log(e^{A} e^{B}) \\
  &= (A + B) + \frac{1}{2}[A,B]
     +\frac{1}{12}\qty([A,[A,B]] + [B,[B, A]]) \notag\\
  &\phantom{={}}+\frac{1}{24} [A,[B,[B,A]]] + \cdots
\label{eq:SchurDynkin} \\
  &\equiv z_1(A,B) + z_2(A,B) + z_3(A,B) + z_4(A,B) + \cdots \\
  &= \sum_\text{odd $n>0$}  z_n(A,B)
    +\sum_\text{even $n>0$} z_n(A,B) \notag\\
  &\equiv z_\text{odd}(A,B) + z_\text{even}(A,B),
\end{align}
\end{subequations}
where the binary product ``$\bLozenge$'' notation is borrowed from Ref.~\onlinecite{Bonfiglioli2012} (p.~115) with $z_1(A,B) = A+B$, $z_2(A,B) = \frac{1}{2}(AB - BA)$, $z_3(A,B) = \frac{1}{12}(A^2B+BA^2 + B^2A+AB^2 - 2ABA - 2BAB)$, $z_4(A,B) = \frac{1}{24}(A^2B^2 - B^2A^2 - 2ABAB + 2BABA)$, etc. The term $z_n(A,B)$ is uniquely expressed in the form of polynomials, but the commutator form is not unique due to Jacobi identity of the Lie algebra. In the above, rearranging the odd-degree and even-degree terms into two sums implies the assumption that the original series is absolutely convergent.

We define the symmetric Jordan--Trotter product form as the symmetrized product (denoted as ``$\circ$'') within an \emph{associative} algebra $(\mathcal{A}, \cdot)$. For any $U, V\in \mathcal{A}$, the not-necessarily-associative, but \emph{commutative}, $\circ$ operation between $U$ and $V$ is defined as
$
    U \circ V 
  \coloneqq \frac{1}{2} \{U, V\} 
   \equiv \frac{1}{2} (UV + VU).
$ 
This defines a \emph{special} Jordan algebra $(\mathcal{A}^{+}, \circ)$. (If the base algebra $\mathcal{A}$ is nonassociative, this defines an exceptional Jordan algebra.) This symmetric Jordan--Trotter product form is a tool to expand the order of a Trotter sequence from $2p$ to $2p+1$, taking the error from $\mathcal{O}(t^{2p+1}) \rightarrow \mathcal{O}(t^{2p+2})$~\cite{chehade2024suzuki}. Specifically, we begin with a first order sequence and analytically eliminate the first order synthesis error. For example, $e^{A+B}$ can be approximated by the symmetric Jordan--Trotter product $e^{A}\circ e^{B} \equiv e^{z_+(A,B)}$, where the generator for the symmetric Jordan--Trotter product is given by
\begin{subequations}
\begin{align}
  &\phantom{={}}   z_+ (A,B)  \notag\\
  &= \log(e^A \circ e^B)
   \equiv \log\qty[\frac{1}{2}\qty(e^{A} e^{B} + e^{B} e^{A})]
\label{eq:z+} \\
  &= (A + B) + \frac{1}{12}([A,[A,B]] + [B,[B,A]]) \notag\\
  &\phantom{={}}   + \qty{ \frac{1}{8}[A,B]^2  + \mathcal{O} (\text{fifth order terms}) }, \label{eq:z+_4th_order}
\end{align}
\end{subequations}
which can be compared with the odd-degree BCH generator
\begin{subequations}
\begin{align}
  &\phantom{={}}   z_\text{odd}(A,B) \notag \\
  &= A \circ_{\bLozenge} B
   \equiv  \frac{1}{2}\qty(A \bLozenge B + B \bLozenge A) \\
  &= \frac{1}{2}\qty[z(A,B) + z(B,A)] \\
  &= \frac{1}{2}\qty[\log(e^A e^B) + \log(e^B e^A)] \\
  &= (A + B) + \frac{1}{12}([A,[A,B]] + [B,[B,A]]) \notag\\
  &\phantom{={}} + \qty{\mathcal{O} (\text{fifth order commutators})} , \label{eq:zOdd_3rd_order}
\end{align}
\end{subequations}
where ``$\circ_{\bLozenge}$'' denotes a special Jordan algebra derived from the $\bLozenge$ product.

\Cref{eq:zOdd_3rd_order} only has odd degree terms, so the first term omitted is the degree-5 term. \Cref{eq:z+_4th_order,eq:zOdd_3rd_order} agree up to the degree-3 terms. In \cref{eq:z+_4th_order}, the coefficient of degree-4 term is not a nested commutator. This is because $[A,B]^2$ does not belong to the free Lie algebra, as indicated by the criterion in the Dynkin--Specht--Wever lemma/theorem~\cite{Dynkin2000, [][{, p145, Lemma 3.26.}]Bonfiglioli2012}. However, for a given representation, such as the defining matrix representation of the Lie algebra, it is possible that $[A,B]^2$ can be written as a Lie algebra element or even nested commutators only involving $A$ and $B$ [consider the Pauli matrix representation of $\mathfrak{su}(2)$ as an example]. Due to the commutativity of the Jordan product, \cref{eq:z+_4th_order} has the inversion symmetry $I z_+(A,B) = z_+(B,A) = z_+(A,B)$, while the TRIS is respected up to the third order, since at fourth order $z_+(A,B) + z_+(-B, -A) \approx \frac{1}{4}[A,B]^2 \neq 0$ for noncommutative $A$ and $B$. 
The difference between \cref{eq:z+_4th_order,eq:zOdd_3rd_order} is enclosed in the curly brackets.
The leading-order difference is in that $z_\text{odd}(A,B)$'s fourth order vanishes exactly. The higher-order terms are also different.


We define the nonunitary time evolution operator by introducing the time parameter explicitly
\begin{align}
     U_{\pm}(t) 
  &= \frac{1}{2} \qty(e^{tA} e^{tB} \pm e^{tB} e^{tA}).
  \label{eq:Ut_pm}
\end{align}
Note that in the generator form, $U_+(t) = e^{z_+(tA, tB)}$, but the analogous $z_-$ generator for $U_{-}(t)$ is not well defined since $U_-$ is not close to identity operator as $t\to 0$. In contrast to the Trotter approximation of the time evolution operator $e^{tH}$ for the Hamiltonian $H = A + B$, the operator $U_+(t)$ preserves the I-symmetry and it also respects TRIS up to $\order{t^4}$. In \cref{tab:symmetry_Qop}, we compare our proposed methods with conventional Trotter product formulas in terms of the TR, I, and TRIS (i.e., TR${}\times{}$I) symmetries.
\begin{table*}[]
  \centering
\begin{tabular}{@{\extracolsep{8pt}}cccccc|c}
\toprule
 \multirowcell{2}[-2pt][c]{Quantum Process\\ $\mathcal{U}(t; A,B)$} &\multicolumn{2}{c}{Time Reversal (TR: $t \leftrightarrow -t$)} &\multicolumn{2}{c}{Inversion (I: $A \leftrightarrow B$)} 
 &\multicolumn{1}{c}{\multirowcell{2}[-2pt][c]{TR${}\times{}$I}}
 \\
\cmidrule{2-3} \cmidrule{4-5}
 &TR symmetry &error $\epsilon_\text{TR}$ &I symmetry &error $\epsilon_\text{I}$ \\
\midrule

 $e^{t(A+B)} \equiv U(t)$
 &\cmark    &0 
 &\cmark    &0
 &\cmark
 &\multirowcell{4}{\rotatebox[origin=c]{90}{\textbf{Case 1}}}
 \\

 $e^{tA} e^{tB} \equiv V(t) \approx U(t)$
 &\xmark    &$\order{t^2}$
 &\xmark    &$\order{t^2}$
 &\cmark
 \\

 $e^{tA/2} e^{tB} e^{tA/2} \approx U(t)$
 &\cmark    &0
 &\xmark    &$\order{t^3}$
 &\xmark
 \\ 

 $\frac{1}{2}(e^{tA} e^{tB} + e^{tB} e^{tA}) \approx U(t)$
 &\xmark    &$\order{t^4}$
 &\cmark    &0
 &\xmark
 \\ 

\midrule

 $e^{t(A+B)} + e^{-t(A+B)}
  = (\mathds{1} + \text{TR})U(t)$
 &$+$: \cmark   &0
 &$+$: \cmark   &0
 &$+$: \cmark   
 &\multirowcell{8}{\rotatebox[origin=c]{90}{\textbf{Case 2}}}
 \\

 $e^{t(A+B)} - e^{-t(A+B)}
  = (\mathds{1} - \text{TR})U(t)$
 &$-$: \cmark   &0
 &$+$: \cmark   &0
 &$-$: \cmark
 \\

 $e^{tA}e^{tB} + e^{-tA}e^{-tB}
  = (\mathds{1} + \text{TR})V(t)$
 &$+$: \cmark   &0
 &\xmark        &$+$: $\order{t^2}$
 &\xmark
 \\

 $e^{tA}e^{tB} - e^{-tA}e^{-tB}
  = (\mathds{1} - \text{TR})V(t)$
 &$-$: \cmark   &0
 &\xmark        &$-$: $\order{t}$
 &\xmark
 \\

 $e^{tA}e^{tB} + e^{-tB}e^{-tA}
  = (\mathds{1} + \text{TR}{\times}\text{I})V(t)$
 &\xmark        &$+$: $\order{t^3}$
 &\xmark        &$+$: $\order{t^3}$
 &$+$: \cmark
 \\

 $e^{tA}e^{tB} - e^{-tB}e^{-tA}
  = (\mathds{1} - \text{TR}{\times}\text{I})V(t)$
 &\xmark        &$-$: $\order{t^2}$
 &\xmark        &$-$: $\order{t}$
 &$-$: \cmark
 \\

 $e^{tA}e^{tB} + e^{tB}e^{tA}
  = (\mathds{1} + \text{I})V(t)$
 &\xmark        &$+$: $\order{t}$
 &$+$: \cmark   &0
 &\xmark
 \\

 $e^{tA}e^{tB} - e^{tB}e^{tA}
  = (\mathds{1} - \text{I})V(t)$
 &\xmark        &$-$: $\order{t^2}$
 &$-$: \cmark   &0
 &\xmark
 \\

\bottomrule
\end{tabular}
\caption{A table of TR ($t \leftrightarrow -t$) and I ($A \leftrightarrow B$) symmetries for various unitary and non-unitary quantum dynamic processes.
$\mathcal{U}(t; A,B)$ in case 1: time evolution unitary $U(t)$ and its approximate formulas by Trotter and Jordan--Trotter product forms.
$\mathcal{U}(t; A,B)$ in case 2: dynamics processes with different symmetries, where $\frac{1}{2}(\mathds{1}\pm \mathcal{S})$ ($\mathcal{S} =$ TR, I, TR${}\times{}$I) are the symmetry ($+$) and anti-symmetry ($-$) projection operators.
In case 1, the TR-symmetry-breaking error $\epsilon_\text{TR} \equiv 
\mathcal{U}(t; A,B) - [\mathcal{U}(-t; A,B)]^{-1}
$, while in case 2, $\epsilon_\text{TR} \equiv \mathcal{U}(t; A,B) - s\mathcal{U}(-t; A,B)$ where $s=\pm$ for TR symmetry and anti-symmetry. In both cases, the I-symmetry-breaking error $\epsilon_\text{I} \equiv \mathcal{U}(t; A,B) - s\mathcal{U}(t; B,A)$ where $s=\pm$ for I symmetry and anti-symmetry.}
\label{tab:symmetry_Qop}
\end{table*}

Our discussion so far has been opaque with regard to the implementation details defining how these operators are synthesized on quantum computers. In \cref{sec:apps}, we consider a concrete implementation and their repercussions regarding BCH-like expansion synthesis.

\subsection{Gadget for exchange-symmetric nonunitary operators} \label{sec:gadget}

We first demonstrate that the gadget in \cref{fig:circuit} gives our symmetrized operator using an ancilla qubit $\ket{q}_a$. Subsequently, we examine the success probability of observing each scattering trajectory of the ancilla qubit $\ket{0}_a \to \ket{0}_a$ ($\ket{1}_a$) in the measurements of the ancilla qubit. These trajectories correspond to the term with the $+$ ($-$) phase factor in front of $V$ in the two block-encoded nonunitary operators $U_\pm \equiv (U \pm V)/2$. The terms are essentially ``kicked back'' from the sign of interference matrix elements $\ev{0 \op{-} 0}_a$ ($\ev{1 \op{-} 0}_a$) of the ancilla qubit (see \cref{sec:gadgetCircuitProof} for how these matrix elements arise). Here, $\ket{\pm}_a = (\ket{0}_a \pm \ket{1}_a)/\sqrt{2}$ are the eigenstates of the $X_a$ operator, and $\ket{+}_a = H_a \ket{0}_a$ and $\ket{-}_a = H_a \ket{1}_a$, where $H_a = (X_a + Z_a)/\sqrt{2}$ is the Hadamard gate acting on the ancilla qubit. In \cref{sec:gadgetCircuitProof}, we show that, given an initial state of $\ket{0}_a\otimes \ket{\Psi}$, the gadget \cref{fig:circuit} block-encodes $U_\pm$ and results in the following scattered final (normalized) state (before the ancilla qubit measurement)
\begin{align}
    \ket{0}_a \otimes U_+\ket{\Psi} 
  + \ket{1}_a \otimes U_-\ket{\Psi}.
  \label{eq:scatteredState_a0}
\end{align} 
During the ancilla qubit measurement, the transition probability $P_{0\to 0}$ ($P_{0\to 1}$) for $\ket{0}_a \to \ket{0}_a$ ($\ket{1}_a$) is given by squared norms of the interference amplitudes: 
\begin{align}
     P_{0\to 0}
  &= \norm{U_+\ket{\Psi}}^2
   = \ev**{\frac{(U^\dagger + V^\dagger)(U+V)}{4}}{\Psi} \\
  &= \frac{1}{2} \qty(1 + \Re\!\ev{V^\dagger U}{\Psi})
   \leq 1,  \\
     P_{0\to 1}
  &= \norm{U_-\ket{\Psi}}^2
   = \ev**{\frac{(U^\dagger - V^\dagger)(U-V)}{4}}{\Psi} \\
  &= \frac{1}{2} \qty(1 - \Re\!\ev{V^\dagger U}{\Psi})
   \leq \frac{1}{4}\norm{U-V}^2.
\end{align}
Since $P_{0\to 0} + P_{0\to 1} = 1$, we have $P_{0\to 0} \geq 1 - \norm{U-V}^2/4$. Consider the unitary operators $U = e^{tA}e^{tB}$ and \mbox{$V = e^{tB}e^{tA}$,} with $V^\dagger = V^{-1} = e^{-tA} e^{-tB}$ and anti-Hermitian operators $A$ and $B$. Then as $t\to 0$, we have 
\begin{align}
  U-V &= [A,B]t^2 + \order{t^3},
\end{align}
with
\begin{align}
  P_{0\to 1} &\leq \norm{U-V}^2/4 \approx \norm{[A,B]}^2 t^4/4
\end{align}
and
\begin{align}
  P_{0\to 0} &\geq 1 - \norm{U-V}^2/4 \approx  1 - \norm{[A,B]}^2 t^4/4.
  \label{eq:P00_t_expansion}
\end{align}

Similarly, given an initial state $\ket{1}_a\otimes \ket{\Psi}$, the scattered final (normalized) state before the ancilla qubit measurement is
\begin{align}
    \ket{0}_a \otimes U_-\ket{\Psi} 
  + \ket{1}_a \otimes U_+\ket{\Psi},
  \label{eq:scatteredState_a1}
\end{align}
and thus we have $P_{1\to 1} = P_{0\to 0}$ and $P_{1\to 0} = P_{0\to 1}$. These ancilla-qubit-state transition probabilities are summarized as follows:
\begin{align}
\begin{tikzpicture}[
 baseline={([yshift=-.5ex]current bounding box.center)},
 node distance=3cm, auto,
 every state/.style={inner sep=2pt,minimum size=0pt},
]
  \node[state] (s0)               {$\ket{0}_a$};
  \node[state] (s1) [right of=s0] {$\ket{1}_a$};
  \path[->] 
    (s0) edge[loop left] node {$P_{0\to 0}$} ()
    (s1) edge[in=15,out=-15,loop right] node {$P_{1\to 1}$} ()
    (s0) edge[bend left] node[below] {$P_{0\to 1}$} (s1)
    (s1) edge[bend left] node[above] {$P_{1\to 0}$} (s0);
\end{tikzpicture}
\label{eq:state_transition_diagram}
\end{align}
Note that these probabilities depend on the input state of principal system $\ket{\Psi}$.

When the gadget is repeatedly applied, the ancilla measurement outcomes $(b_1, b_2, \dots, b_\kappa, \dots)$, with  $b_\kappa \in \{0, 1\}$, constitute a time-varying Markov chain with the time-dependent transition probability matrix $\Gamma_\kappa$ (left stochastic matrix convention),
\begin{align}
  &  \Prob(b_{\kappa} \to b_{\kappa+1}) 
  \equiv \Gamma_\kappa 
   = \pmqty{P_{0_{\kappa} \to 0_{\kappa+1}} & P_{1_{\kappa} \to 0_{\kappa+1}} \\ 
            P_{0_{\kappa} \to 1_{\kappa+1}} & P_{1_{\kappa} \to 1_{\kappa+1}}}, \\
  &  P_{0_{\kappa} \to 0_{\kappa+1}} 
   = P_{1_{\kappa} \to 1_{\kappa+1}}
   = \norm{U_{+} \ket{\Psi_\kappa}}^2 \equiv P_{\kappa +}, \label{eq:tranProb+}\\
  &  P_{0_{\kappa} \to 1_{\kappa+1}} 
   = P_{1_{\kappa} \to 0_{\kappa+1}}
   = \norm{U_{-} \ket{\Psi_\kappa}}^2 \equiv P_{\kappa -} .
\end{align}
The $\Gamma$ is also known as the rate matrix (hence the use of the symbol $\Gamma$). The quantum states $(\ket{\Psi_1}, \ket{\Psi_2}, \dots, \ket{\Psi_\kappa}, \dots)$ essentially also constitute a Markov chain with state transitions $\ket{\Psi_{\kappa+1}} = U_{\pm} \ket{\Psi_{\kappa}}/\sqrt{P_{\kappa \pm}}$ and corresponding transition probabilities $P_{\kappa \pm}$. Combined with $(b_1, b_2, \dots, b_\kappa, \dots)$, our system becomes a hidden quantum Markov model~\cite{Monras2011,Milz2021}. These definitions and stochastic analysis will be used in \cref{sec:apps-Qwalk}. For example, the bit matrix plotted in \cref{fig:RWsimulations}(b) represent the emission states $b$-vector here and the estimated transition probabilities can be used for spectroscopy analogous to quantum phase estimation.

As a side note, instead of the (anti-)symmetric form, a general LCU of two unitary operators, $U$ and $V$, can be constructed by replacing the first (second) Hadamard gate with a general unitary operator on the ancilla qubit $\hat{G} = \sum_{a,b \in \{0,1\}} g_{ab} \op{a}{b}$ ($\hat{G}^\dagger$). We obtain similar results as follows.
\begin{align}
    \ket{0}_a \otimes \ket{\Psi} \mapsto 
  & \ket{0}_a \otimes \qty[ |g_{00}|^2 U + (1-|g_{00}|^2) V] \ket{\Psi} + \notag \\
  & \ket{1}_a \otimes \qty[ g_{00}^{\vphantom{*}} g_{01}^* (U - V)] \ket{\Psi}, \\
    \ket{1}_a \otimes \ket{\Psi} \mapsto 
  & \ket{1}_a \otimes \qty[ |g_{01}|^2 U + (1-|g_{01}|^2) V] \ket{\Psi} + \notag \\
  & \ket{0}_a \otimes \qty[ g_{00}^* g_{01}^{\vphantom{*}} (U - V)] \ket{\Psi}, \\
     P_{0\to 1} = P_{1\to 0} 
  &= 2|g_{00}|^2 |g_{01}|^2\qty(1 - \Re\!\ev{V^\dagger U}{\Psi}), \\
     P_{0\to 0} = P_{1\to 1} 
  &= 1 - P_{0\to 1}.
\end{align}
In the above, we used the unitary property $\hat{G}^\dagger \hat{G} = \mathds{1}$. Specifically, $|g_{00}|^2 + |g_{10}|^2 = |g_{01}|^2 + |g_{11}|^2 = 1$ and $g_{00}^{\vphantom{*}} g_{01}^* + g_{10}^{\vphantom{*}} g_{11}^* = g_{01}^{\vphantom{*}} g_{00}^* + g_{11}^{\vphantom{*}} g_{10}^* = 0$. Since $\hat{G}\hat{G}^\dagger = \mathds{1}$ is also true, it follows that $|g_{00}|^2 + |g_{01}|^2 = 1$. Additionally, we obtain $P_{0\to 1} = P_{1\to 0} \leq \frac{1}{2}(|g_{00}|^2 + |g_{01}|^2)^2 (1 - \Re\!\ev{V^\dagger U}{\Psi}) = \frac{1}{2} (1 - \Re\!\ev{V^\dagger U}{\Psi})$ for any $\hat{G}$, with the maximal value achieved whenever $|g_{00}| = |g_{01}| = |g_{10}| = |g_{11}| = 1/\sqrt{2}$. In particular, the Hadamard gate case corresponds to $(g_{00}, g_{01}, g_{10}, g_{11}) = (1, 1, 1, -1) / \sqrt{2}$ with a maximal $P_{0\to 1} = P_{1\to 0}$ [corresponding to odd-inversion symmetric evolution $\frac{1}{2}(U-V)$], hence a minimal $P_{0\to 0} = P_{1\to 1}$ [corresponding to even-inversion symmetric evolution $\frac{1}{2}(U+V)$], for any fixed $U$, $V$, and $\ket{\Psi}$.

\section{Applications} \label{sec:apps}
\subsection{TRIS and TRIS-Breaking Spectral Walks} \label{sec:apps-Qwalk}

In addition to the quantum simulation of complex objects, quantum computers offer an avenue to probe fundamental concepts such as the emergence or breaking of a symmetry \cite{lacroix2023}. Time-reversal is one such symmetry whose presence has fundamental consequences for the resulting spectra (Kramers' theorem) and dynamics. An operator is time-reversal symmetric if it is invariant under $t \rightarrow -t$. Time reversal symmetry is often tied to or discussed in conjunction with parity inversion symmetry, e.g., in the well-known CPT symmetry. As mentioned earlier, the I-symmetry in TRIS of the Jordan product coincides with the parity symmetry for a bipartite lattice. 

We use this fact and the Jordan product to define an algorithm for TRIS and TRIS-breaking quantum random walks. Motivated by near-term resource limitations, we focus on the stabilization of a TRIS random walk. Next, we apply the TRIS random walk to spectroscopy and spectral projection. Inspired by prior work, relating random walks to spectral identification~\cite{Chen2020, Granade2022, Ferris2023}, we remark on how our algorithm could be utilized for spectral projection in a unitary and non-unitary setting.  Finally, we discuss the extent to which random coherent imprecision can break the time reversal symmetry. Steps to recover TRIS are discussed as well.

\subsubsection{Analysis of a single step} \label{sec:randomWalk_1step}
Each step of the random walks we will consider is implemented with the symmetry gadget in \cref{fig:circuit} as follows. Consider the Jordan forms as the linear operators comprised of forward and backward time evolutions:
\begin{align*}
    U_+(t) &= \frac{U(t) + U(-t)}{2} = \frac{e^{-iHt} + e^{iHt}}{2} = + \cos(Ht), \\
    U_-(t) &= \frac{U(t) - U(-t)}{2} = \frac{e^{-iHt} - e^{iHt}}{2} = -i\sin(Ht). 
\end{align*}
In this special example, where $V = [U(t)]^\dagger = U(-t)$, the time-reversed evolutions interfere such that the overlap factors becomes ${\Re}\ev{V^\dagger U}{\Psi} = {\Re}\ev{U^2}{\Psi} = \ev{\cos2Ht}{\Psi}\equiv\cos{2Et}$. Here, $E$ is a function of $\ket{\Psi}$. For small $|t|$, an approximate solution to $E$ is given by $E\approx \ev{H^2}{\Psi}^{1/2}$. The ancilla transition probabilities then encode the trigonometric power reduction relations, namely, $P_{0\to 0} = (1+ \cos{2Et})/2 = \cos^2{Et}$ and similarly, $ P_{0\to 1}=(1 - \cos{2Et})/2 = \sin^2{Et}$.

\subsubsection{Spectroscopy by many-shot random walks} \label{sec:decoherence}
From an initial state $\ket{0}_a\otimes \ket{\Psi}$, after the first-step walk and measurement of the ancilla, the principal system evolves into the normalized state $\ket{\Psi_1} = \cos(Ht)\ket{\Psi}/\sqrt{P_{0\to 0}}$ [$-i\sin(Ht)\ket{\Psi}/\sqrt{P_{0\to 1}}$] with probability $P_{0\to 0}$ ($P_{0\to 1}$). If we run the experiments with many shots, each shot being a single-step walk with time-step size $t$, we will obtain a mixed state density operator $\rho(t)$ from a pure initial state $\rho(0) \equiv \rho \equiv \op{\Psi}$. $\rho(t)$ is given by
\begin{align}
\label{eq:csHt}
     \rho(t) 
  &= \cos(Ht)\rho\cos(Ht) + \sin(Ht)\rho\sin(Ht) \\
  &\equiv  \mathcal{K}_{0,t}(\rho) + \mathcal{K}_{1,t}(\rho) 
  \equiv \mathcal{H}^\text{cs}_t(\rho),
\end{align}
which can be used to describe the measurements of an observable $\ev*{\hat{O}} = {\Tr}(\rho(t)\hat{O})$ on the principal system. The evolution above is nonunitary, but rather described by a Lindblad evolution process, as a sum of Kraus maps $\mathcal{K}_{0,t}(\rho) = \cos(Ht)\rho\cos(Ht)$ and $\mathcal{K}_{1,t}(\rho) = \sin(Ht)\rho\sin(Ht)$.
Note that our time evolution channel given by \cref{eq:csHt} is time reversal symmetric if we do not post-select the ancilla measurement outcome strings. In addition, it conserves the energy $\Tr(\rho(t) H ) = \Tr (\rho(0)H) = \ev{H}{\Psi}$.

Since the density operator under unitary evolution channel is
\begin{align}
\label{eq:unitaryHt}
     \rho_U (t) 
  &= \mathcal{H}_t(\rho) = e^{-iHt}\rho e^{iHt} \\
  &= \cos(Ht)\rho\cos(Ht) + \sin(Ht)\rho\sin(Ht) \notag\\
  &\mathrel{\phantom{=}} +i\cos(Ht)\rho\sin(Ht) -i\sin(Ht)\rho\cos(Ht),
\end{align}
we have $\rho(t) = [\rho_U(t) + \rho_U(-t)]/2$, i.e., $\mathcal{H}^\text{cs}_t(\rho) = (\mathcal{H}_t + \mathcal{H}_{-t})/2$. That is, our time-symmetric channel $\mathcal{H}^\text{cs}_t(\rho)$ is equivalent to a random unitary channel that is the sum of $\mathcal{H}_t$ and $\mathcal{H}_{-t}$, resulting a similar circuit used by Ref.~\onlinecite{Granade2022} to perform iterative phase estimation, given an eigenstate of the principal system. However, with the same circuit here, the $t$-sampling algorithm proposed in Ref.~\onlinecite{Choi2021} (parallel version with many ancilla qubits) and Ref.~\onlinecite{Stetcu2023} (iterative version most similar to our method) can be used to perform spectroscopy and spectral projection.

However, by symmetrically sampling $\mathcal{H}_t$ and $\mathcal{H}_{-t}$ following a \emph{multi-step} random walk of ancilla qubit state, the principal system is driven to a stable fixed point that is a diagonal ensemble in the energy eigenbasis of the Hamiltonian $H$. This allows an improved, second, spectroscopy method by simply measuring the Hamiltonian in the diagonal ensemble. 
When compared to another spectral estimation method by Ferris \textit{et al.}~\cite{Ferris2023}, our method does not require the maximally mixed state as a resource or use the noise-sensitive Hadamard test.

To prove that diagonal ensemble is the stable fixed point of our spectral random walk dynamics when the symmetric channel $\mathcal{H}^\text{cs}_t$ is iterated for $r$ steps as $(\mathcal{H}^\text{cs}_t)^{n}(\rho)$, we can use a similar technique given in Ref.~\onlinecite{Novotny2009} by using $p_1=1/2=p_2$ and $U_1= U(t), U_2 = U_1^\dagger = U(-t)$. A more transparent proof without using the monotonicity of von Neumann entropy under the channel $\mathcal{H}^\text{cs}_t$ is as follows.

Denote the energy eigenbasis of a Hermitian Hamiltonian $H$ as $\{(\omega_n, \ket{n})\}$, where $H\ket{n} = \omega_n \ket{n}$ and $\omega_n \in \mbb{R}$, and the initial pure state $\ket{\Psi} = \sum_n c_n \ket{n}$ with the corresponding density matrix $\rho = \sum_{n,m} c_n c^*_m \op{n}{m} = \sum_{k} |c_k|^2 \op{k} + \sum_{n\neq m} c_n c^*_m \op{n}{m} \equiv \rho^\text{d} + \rho^\text{od}$, where $\rho_\text{d}$ ($\rho_\text{od}$) is the diagonal (off-diagonal) part of the density operator in the energy eigenbasis. $\rho_\text{d}$ is a diagonal ensemble. We can assume that there is no degeneracy in $\rho^\text{od}$ part, i.e., $\omega_n \neq \omega_m$ for $n\neq m$.\footnote{Should there be any degeneracy, a unitary transformation within the degenerate subspace can put those terms into the $\rho^\text{d}$ part, while all new basis states are still energy eigenstates. In the case of pure-state density operator $\op{\Psi}$ with $\ket{\Psi} = \sum_n c_n \ket{n}$, this transformation can be simply done by grouping all degenerate eigenstates corresponding to the same eigenenergy into a new eigenstate with a proper normalization.}
The operator $\rho^\text{d}$ does not evolve with time since $\mathcal{H}^\text{cs}_t(\rho^\text{d}) = \mathcal{H}_t(\rho^\text{d}) = \rho^\text{d}$. In other words, $\rho^\text{d}$ is a fixed point for both $\mathcal{H}^\text{cs}_t$ and $\mathcal{H}_t$. However, the corresponding stability of this fixed point under the iterations of $\mathcal{H}^\text{cs}_t$ and $\mathcal{H}_t$ is different. To see this, for a short time $|t|$, comparing
\begin{align*}
     \rho_U(t) 
  &= \rho - it[H, \rho] - \frac{t^2}{2}[H,[H,\rho]] + \order{t^3} \\
  &= \rho^\text{d} + \rho^\text{od} - it[H, \rho^\text{od}] - \frac{t^2}{2}[H,[H,\rho^\text{od}]] + \order{t^3}, \\
\end{align*}
with 
\begin{align*}
     \rho(t)
  &= \frac{\rho_U(t) + \rho_U(-t)}{2}
   = \rho - \frac{t^2}{2}[H,[H,\rho]] + \order{t^4} \\
  &= \rho^\text{d} + \rho^\text{od} - 
     \frac{t^2}{2}[H,[H,\rho^\text{od}]] + \order{t^4},
\end{align*}
and using
\begin{align}
     & \rho^\text{od} - it[H, \rho^\text{od}] \notag\\
  ={}& \sum_{n\neq m} \qty[1 - it(\omega_n - \omega_m)] c_n c^*_m \op{n}{m}, \\
     & \rho^\text{od} - \frac{t^2}{2}[H, [H, \rho^\text{od}]] \notag\\
  ={}& \sum_{n\neq m} \qty[1 - \frac{t^2(\omega_n - \omega_m)^2}{2}] c_n c^*_m \op{n}{m},
\end{align}
we find that the off-diagonal part of time-evolved density operator $\rho(t)$ contracting but not the off-diagonal part of $\rho_U(t)$ due to the first-order term. Therefore, if we apply the channel $\mathcal{H}^\text{cs}_t$ iteratively as
\begin{align}
    \prod_{\kappa=1}^{r} \mathcal{H}^\text{cs}_{t_\kappa}
  = \prod_{\kappa=1}^{r} (\mathcal{K}_{0, t_\kappa} + \mathcal{K}_{1, t_\kappa}),
  \label{eq:r-step_channel}
\end{align}
then the final density matrix will contract to the fixed point $\rho^\text{d} = \sum_k |c_k|^2 \op{k}$ without any remaining interference terms from $\rho^\text{od}$. In \cref{sec:CV_decoherence}, we generalize this decoherence proof to the gadget circuit in \cref{fig:circuit} with a continuous variable ancilla qumode, demonstrating potentially broader application of our algorithm.

We conclude that by running many shots and adding all $r$-step paths (i.e., trajectories) incoherently, our spectral random walk drives the principal system to a \emph{stationary} mixed state $\rho = \sum_k |c_k|^2 \op{k}$ that is diagonal in the energy eigenbasis after passing the \emph{mixing time}~\cite{Levin2017}. The population of the energy eigenstates is given by the Born's rule probabilities $|c_k|^2$, which is determined by the initial pure state $\ket{\Psi} = \sum_k c_k \ket{k}$.

\subsubsection{Spectral projection by a single-shot semicoherent random walk} \label{sec:randomWalk}

\begin{figure*}[tb]
 \centering
 \includegraphics[scale=1]{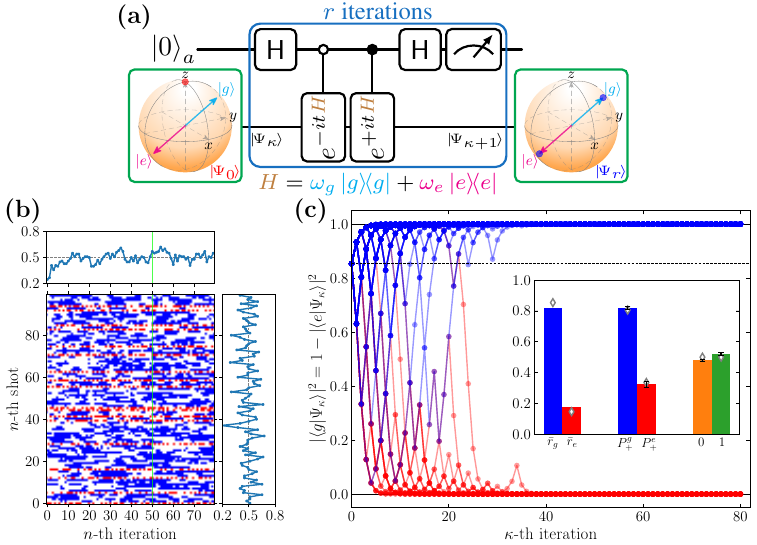}
\caption{(a) Spectral projection circuit derived from symmetry gadget in \cref{fig:circuit}. The system initial state $\ket{\Psi}=\ket{0}$ is indicated by the red dot at the north pole of the left Bloch sphere. The evolved final state (blue dots on the right Bloch sphere) is projected onto one of two eigenstates, the ground state $\ket{g}$ and the excited state $\ket{e}$, of the two-level system Hamiltonian $H$.
(b) The bit matrix of the ancilla qubit measurement outcome: $0$ (white pixels) or $1$ (red or blue pixels). Red (blue) indicates the random walk path (each row of the bit matrix) reaches to the excited state $\ket{e}$ (ground state $\ket{g}$) within $r=80$ iterations (measurements). The blue curve of right inset (top inset) is the ratio of measurement outcome $1$ for each path (for each iteration step over all shots).
(c) The system qubit's fidelity with respect to the ground state $\ket{g}$ at each iteration step after each measurement of the ancilla qubit. The red (blue) indicates the excited (ground) state fidelity reaches to 1 within $r$ iterations.
Inset of (c): First two bars are ratio of ground state (blue) and excited state (red) paths.
Next two bars are estimated $0\to 0$ or $1\to 1$ transition probabilities defined in \cref{eq:tranProb+} and calculated using the bit matrix elements on the right of the vertical green line in panel (b) when the eigenstates have been projected out and the Markov chains become time-independent. Due to the same-color pixel transition $P_+^g > 0.5 > P_+^e$, the excited state paths (red) are more fragmented than ground state paths (blue).
Last two bars are the $0$ and $1$ ratios after $50$ iterations (``warm-up'' period). The green bar can be viewed as the mean values of the data after $50$ iterations in the top inset of panel (b).
The diamond symbols in the inset of (c) are exact values of all estimates.}
 \label{fig:RWsimulations}
\end{figure*}

With a same initial state $\ket{\Psi}$, the diagonal ensembles $\rho^\text{d}$'s obtained with the equivalent channels, $\mathcal{H}^\text{cs}_t(\rho)$ and $(\mathcal{H}_t + \mathcal{H}_{-t})/2$, are indistinguishable theoretically. However, the gadget circuit in \cref{fig:circuit} implementing $\mathcal{H}^\text{cs}_t(\rho)$ uniquely unravel the trajectories: each path can be labeled by a bit vector $\vb{b} = (b_1, b_2, \dots, b_r)$ consisting of the sequence of ancilla measurement outcomes $b_\kappa \in \{0, 1\}$. For this path, the initial pure state $\rho = \op{\Psi}$ is evolved by the single term $\prod_{\kappa=1}^{r} \mathcal{K}_{b_\kappa, t_\kappa}\equiv \mathcal{K}_{\vb{b}, \vb{t}}$ from the expansion of the series product \cref{eq:r-step_channel}. The specific form for $\mathcal{K}_{\vb{b},\vb{t}}$ is $\mathcal{K}_{\vb{b}, \vb{t}} (\rho) = (K_{b_r,t_r} \cdots K_{b_1,t_1}) \rho (K^\dagger_{b_1,t_1} \cdots K^\dagger_{b_r,t_r}) \equiv K_{\vb{b},\vb{t}} \rho K^\dagger_{\vb{b},\vb{t}}$, where $K_{0,t} = \cos(Ht)$ and $K_{1,t} = \sin(Ht)$. The evolved final state is the pure state $\rho_{\vb{b},\vb{t}} = \mathcal{K}_{\vb{b}, \vb{t}}(\rho) / P(\vb{b},\vb{t})$, where $P(\vb{b},\vb{t}) = \Tr[\mathcal{K}_{\vb{b}, \vb{t}}(\rho)]$. 

For convenience, we suppress the $t_\kappa$ and $\vb{t}$ subscripts. For sufficiently large total number of steps $r$ and $\vb{b} \in B \subset \{0,1\}^{r}$, we have observed that the evolved final pure state $\rho^{\text{d}}_{\vb{b}} = \op{n_{\vb{b}}}$, where $\ket{n_{\vb{b}}}$ is one of the energy eigenstates, which behaves as absorbing states of our spectral walk. Thus, we can apply random walk to project out all eigenstates contained in the initial state.
This is a stronger result than the mixture of trajectories flowing toward the diagonal ensemble contained in the initial state.

Numerical experiments indicate that the total number of paths absorbed within $r$ steps by the energy eigenstates contained in the initial state $\ket{\Psi} = \sum_n c_n \ket{n}$ is $|B| \approx |\{0,1\}^{r}| = 2^r$, so $\sum_{\vb{b}\in B} P(\vb{b}) \approx \sum_{\vb{b}} P(\vb{b}) = 1$. Since the sum of all $2^r$ paths recovers $ \sum_{\vb{b}}P(\vb{b}) \rho_{\vb{b}} = \sum_{\vb{b}} \mathcal{K}_{\vb{b}}(\rho) = \rho^\text{d} = \sum_n |c_n|^2 \op{n}$ according to \cref{sec:decoherence}, and $\sum_{\vb{b}} P(\vb{b}) \rho_{\vb{b}} \approx \sum_{\vb{b}\in B} P(\vb{b}) \op{n_{\vb{b}}} = \sum_{n} [\sum_{\vb{b}\in B_n} P(\vb{b})]\op{n}$, where $\bigcup_{n} B_n = B$, we conclude that there should be $\approx |B| |c_n|^2$ paths that project out the energy eigenstate $\ket{k}$, or equivalently, the probability $P(\ket{n} \mid \vb{b}\in B) = \sum_{\vb{b}\in B_k} P(\vb{b}) \approx |c_n|^2$, following the Born rule.

While the outcome of our spectral projection, including the Born rule for the probability of projecting certain energy eigenstate contained in the initial state, is similar to that of Ref.~\onlinecite{Chen2020}, our symmetry gadget is simpler than the primitive circuit in Ref.~\onlinecite{Chen2020}: it has fewer free parameters that need to be optimized and the time reversal symmetry enables easier analysis of the spectral walk dynamics. To demonstrate this, we consider a simple, illuminating example with one-qubit system coupled to an ancilla qubit as shown in \cref{fig:RWsimulations}(a). For the system qubit, the two-level system Hamiltonian $H = \omega_g \op{g} + \omega_e \op{e}$, with the ground state $\ket{g}$ and the excited state $\ket{e}$ and by definition $\omega_g \leq \omega_e$. We assume $\omega_g \neq \pm \omega_e$ since the iterative spectral projection fails at these two points: degenerate point $\omega_g = \omega_e$ and particle-hole symmetric point $\omega_g = -\omega_e$. The former case need not prepare eigenstate since any state is one; the latter case can be fixed by adding a constant shift to the Hamiltonian.

Consider the system initial state $\ket{\Psi} = \ket{0} = (z\ket{g} + \ket{e})/\sqrt{|z|^2 + 1}$, $z\in \mathbb{C}$~\cite{Kalman2018}. The effect of the even (odd) time nonunitary evolution $K_{0,t} = \cos(Ht)$ and $K_{1,t} = \sin(Ht)$ can be viewed as M\"{o}bius transformations~\cite{Kalman2018} $\mu(z) = k_{0,1}z$ so that the new normalized quantum state is $\ket{\Psi_1} = (\mu(z)\ket{g} + \ket{e})/\sqrt{|\mu(z)|^2 + 1}$. $k_0 = \cos(\omega_g t)/\cos(\omega_e t)$ for $K_{0,t} = \cos(Ht)$ and $k_1 = \sin(\omega_g t)/\sin(\omega_e t)$ for $K_{1,t} = \sin(Ht)$. These two M\"{o}bius transformations have a common pair of fixed points $z=0$ ($\ket{\Psi} = \ket{e}$) and $z=\infty$ ($\ket{\Psi} = \ket{g}$), but the repulsive and attractive fix points switch since if $|k_0| > 1$, then $|k_1| < 1$ and vice versa. Therefore the convergence is not guaranteed~\cite{Ambroladze2000,McCarthy2018}. However, because the probabilities $P_{\kappa+}$ and $P_{\kappa-}$ also depend on the state $\ket{\Psi_{\kappa-1}}$, for sufficiently large $r$, the transform tends to converge to one of the two fixed points and thus projects out the eigenstates. We have numerically verified this and the results using noiseless quantum circuit simulator are plotted in \cref{fig:RWsimulations} explained in more details as follows. 

The one-qubit Hamiltonian used in the numerical simulation is given by 
\begin{align}
     H 
  &= \omega_g \op{g} + \omega_e \op{e}
   = \omega_{+} \mathds{1} + 
     \omega_{-} \vb{n} \vdot \hat{\vb*{\sigma}},
  \label{eq:RW_Ham_1q}
\end{align}
where $\op{g} = \frac{1}{2}(\mathds{1} + \vb{n} \vdot \hat{\vb*{\sigma}})$, $\op{e} = \frac{1}{2}(\mathds{1} -\vb{n} \vdot \hat{\vb*{\sigma}})$, $\omega_\pm = \frac{1}{2}(\omega_g \pm \omega_e)$, and $\pm\hat{\vb{n}} = \pm (\sin\theta \cos\phi, \sin\theta \sin\phi, \cos\theta)$ are the Bloch vectors for the eigenstates $\ket{g}$ and $\ket{e}$, respectively. Here, the Pauli vector $\hat{\vb*{\sigma}} = (X , Y, Z)$ consists of three components of the Pauli matrices. \Cref{eq:RW_Ham_1q} is the same as \cref{eq:Ham_1q}, but here we use Bloch vector representation to match the representation of the eigenstates on Bloch sphere in \cref{fig:RWsimulations}(a).
In \cref{sec:appRWcirc} we provide more details and the compiled circuit suitable for noisy intermediate-scale quantum (NISQ) computers.
Without loss of generality, we choose $\omega_+ = \sqrt{7}$, $\omega_- = -\sqrt{3}$, $\hat{\vb{n}} = (\frac{1}{2}, \frac{1}{2}, \frac{1}{\sqrt{2}})$ (for $\theta = \phi = \frac{\pi}{4}$), and the time step size $t = 0.5$. In \cref{fig:RWsimulations}(b) and \cref{fig:RWsimulations}(c) we plot the results for $r=80$ iterations (i.e., random walk steps in a path) and $n_s = 100$ shots (random walk paths). Markedly, all paths either project on $\ket{g}$ or $\ket{e}$ and, as shown in the inset of \cref{fig:RWsimulations}(c), the probabilities $\bar{r}_g \approx \cos^2(\theta/2)$ and $\bar{r}_e \approx \sin^2(\theta/2)$ following the Born rule prediction for the initially prepared state $\ket{\Psi_0} = \cos(\theta/2)\ket{g} + \sin(\theta/2)\ket{e}$, which agrees with the observation in Ref.~\onlinecite{Chen2020} that used a less symmetric spectral projection circuit. However, our symmetric circuit structure and simpler gates on ancilla qubit allows a clearer analysis of the quantum process and, in addition to eigenstate projections, the ability to directly estimate the spectrum (e.g., $\omega_\pm$ or $\omega_{g,e}$) with ancilla qubit state transition probabilities.

\subsubsection{Timing of spectral-projection random walk}

To efficiently project out the eigenenergy spectrum, we need to choose proper time step sizes $\{t_\kappa \}_{\kappa=1}^r$. First, for very short time evolution step sizes, $P_{0\to 0} \gg P_{0\to 1}$. In this case, consecutive $\cos(Ht)$ operators are applied to the state with high probability and the resulting operator $\cos^r(Ht) = e^{r\log[\cos(Ht)]} \approx e^{-rH^2t^2/2}$ projects out the ground state (assume the spectrum is shifted to $\geq 0$ and there is finite overlap between the ground state and initial state \cite{Ge2019, Lin2020, Keen2021}). Therefore, we can apply this algorithm to prepare ground state by frequently measuring the ancilla. However, too frequent measurements would result in nonevolving initial state due to quantum Zeno effect (QZE). If this happens, we would have a quadratically stronger QZE (i.e., slower evolution) due to the time evolution operator $\cos(Ht)$ instead of the usual $e^{-iHt}$. We check this in \cref{sec:appQZE} by computing $\lim_{n\to \infty} S_n(T/n)$, conditioned that the ancilla measurement is always $0$, where the survival probability $S_n(T/n) = \abs{\ip{\Psi}{\Phi}}^2$ and the time evolved final state is $\ket{\Phi} = \cos^n(HT/n)\ket{\Psi}/ [\ev{\cos^{2n}(HT/n)}{\Psi}]^{1/2}$. A subtle difference with a textbook QZE is that the principal system can be kept stationary by measuring an ancillary system that couples to the principal system. In a textbook example, the principal system is directly measured. This provides an advantage to use our setting to observe or achieve the QZE since measuring a single ancilla qubit is much simpler than measuring the entire principal system.

To find a suitable time step sizes for spectral projection, we rewrite \cref{eq:csHt} in the energy eigenbasis
\begin{align}
     \rho(t) 
  &= \rho^\text{d} + \sum_{n\neq m} [\cos(\omega_n t) \cos(\omega_m t) \notag\\
  &\mathrel{\phantom{=}}  +{} \sin(\omega_n t) \sin(\omega_m t)] c_n c^*_m \op{n}{m} \notag\\
  &= \rho^\text{d} + \sum_{n\neq m} \cos[(\omega_n - \omega_m )t] c_n c^*_m \op{n}{m}.
\end{align}
After $r$ steps, the coefficient for $\op{n}{m}$ is suppressed by a factor $\prod_{\kappa=1}^{r} \cos[(\omega_n - \omega_m)t_\kappa]$. Since $|\cos[(\omega_n - \omega_m)t_\kappa]| \leq 1$ and $\omega_n \neq \omega_m$ for $n\neq m$, the projection will always succeed for sufficiently large $r$ and any set of $t_\kappa$'s [$\not \equiv 0 \pmod{\pi/|\omega_n - \omega_m|}$] larger than the QZE time scale set by $1/r$.

\subsubsection{Symmetric spectral walks}

We start from an initial state $\ket{\Psi_0}$. At each clock-cycle $\kappa \in \{1, 2, \dots, r\}$, select a random variable $t_\kappa$. Using an ancilla qubit and evolve according to the evolution $\frac{1}{2}\qty[e^{-iHt_\kappa} + (-1)^{b_\kappa} e^{iHt_\kappa}]$, where $b_\kappa = m_{\kappa-1} \oplus m_{\kappa}$ without resetting the ancilla, $b_\kappa = 0 \oplus m_{\kappa}$ with resetting the ancilla to $\ket{0}_a$, and $m_\kappa \in \{0,1\}$ is the measurement outcome obtained at the end of this cycle $\kappa$. After $r$ clock-cycles, the state has evolved from $\ket{\Psi_0}$ to $\ket{\Psi_r} = F_r \ket{\Psi_0}$ (not normalized) according to  $F_r = \prod_{\kappa=1}^{\kappa = r} f_{a_\kappa}(H t_\kappa)$, where $f_{a_\kappa = 0}(\cdot) = \cos(\cdot)$, occurring with probability as above $P_{0\to 0}=\cos^2{E_{r-1}t_r}$. Similarly, $f_{a_\kappa = 1} (\cdot)= -i\sin(\cdot)$ with probability $P_{0\to 1}=\sin^2{E_{r-1}t_r}$, where $\cos2E_{r-1}t_r = \ev{\cos2Ht_r}{\Psi_{r-1}}/\ip{\Psi_{r-1}} = \ev{(\cos2Ht_r)F_{r-1}^\dagger F_{r-1}}{\Psi_{0}}/\ev{F_{r-1}^\dagger F_{r-1}}{\Psi_0}$ (note that $[H, F_r] = 0$). 

Since the relative ratio of $P_{0\to 0}/P_{0\to 1} = \cot^2{Et}$, it follows that the ancilla qubit realizes a heavily biased random coin producing many strings of $a_\kappa = 0$ interspersed with the odd counter driving term $-i\sin{Ht}$. TRIS-preserving operator strings reduce to $\cos^\kappa_{Q} (Ht)$, which can be expanded into a binomial expansion using the trig identity $\cos^2 Ht = (\mathds{1}+\cos(2Ht))/2$. Again, using our prior trick, $\sin^2 Ht = (\mathds{1}-\cos(2Ht))/2$, the \textit{product} of two TRIS-breaking terms \textit{preserves} the TRIS. While actions at different times have opposite temporal parities, they nevertheless commute so that $\prod_{\kappa=1}^{r} f_{a_\kappa}(H t_\kappa) = \prod_{\kappa \in \mathcal{I}_0}\cos (H t_\kappa) \times (-i)^{|\mathcal{I}_1|}\prod_{\kappa\in \mathcal{I}_1}\sin (H t_\kappa)$, where $\mathcal{I}_0 = \{\kappa \in [r] \mid a_\kappa = 0\}$, and $\mathcal{I}_1 = \{\kappa \in [r] \mid a_\kappa = 1\}$. That is, any configuration with an \textit{even} number (denoted as $|\mathcal{I}_1|$) of $0\to1$ outcomes is a valid TRIS coherent configuration, while those with $|\mathcal{I}_1|$ \textit{odd} break the TRIS. It is easy to see that breaking TRIS and breaking the Hermiticity of the $F_r$ operator occur in tandem. 

Additionally, the times $t_\kappa$ can be adjusted to explore different TRIS distributions. The general expression can likewise be factorized using $\cos Ht \cos H \tau  = \{\cos[H(t-\tau)]+\cos[H(t+\tau)]\}/2$. In the construction above, we constructed an operator $\propto U(t) + U(-t).$ Let us generically consider a case where a control error $2 \delta$ is present in the form:
\begin{align*}
  &\mathrel{\phantom{=}} 
     U(t) + U(-t+2\delta)  \\
  &= [U(t-\delta ) + U(-t+\delta)] U(\delta)  \\
  &= 2\cos[H(t-\delta)] [\cos(H\delta) +i \sin(H\delta)].
\end{align*}
As the backwards time-evolution was shortened by a factor $2\delta$, we interpret the expression above as a transformation to a center-of-time coordinate, namely $t_c = t-\delta$ and a relative-time coordinate $\delta$. The first term, which is a product of cosines, is time-reversal symmetric while the second term, involving the sine, breaks the TRIS. 

\subsection{Trotter Factorized Spin Procession} \label{sec:apps-TrotterFactorization}

\begin{SCfigure*}
 \centering
 \includegraphics[scale=0.85]{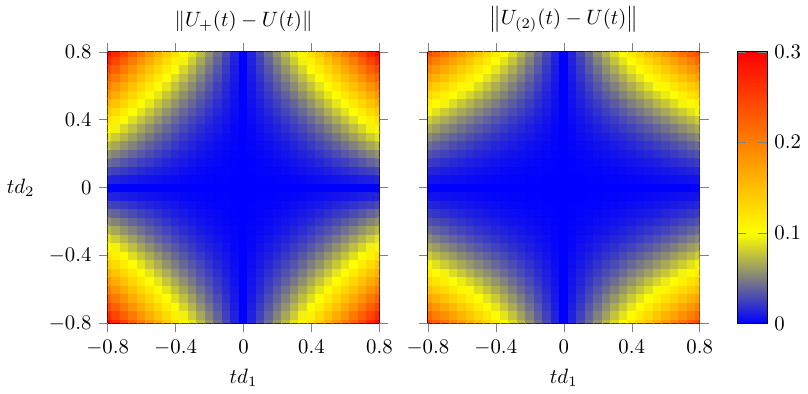}
\caption{Comparison of the operator norm error between two-unitary LCTU $U_+(t)$ and the second order Trotter $U_{(2)}(t)$ in approximating the exact unitary $U(t)$. Generally, $U_{(2)}$ has smaller error than $U_+$. The error plot's symmetry can be understood from their respective small $t$ expansion: $U_+(t)$ has fourfold symmetry due to $|\sin(2\theta)|$ factor, while $U_{(2)}(t)$ only has twofold symmetry due to additional $\sqrt{5-3\cos(2\theta)}$ factor.}
  \label{fig:1qTrotterError}
\end{SCfigure*}

In order to illustrate and numerically confirm the validity of \cref{eq:Ut_pm}, we consider a minimal model of semicoherent Trotter workflow.

The simplest illustrative model of quantum computation is a single qubit system. This is because its true dynamics and the approximate Trotter dynamics~\cite{Hatano2005} can both be exactly solved on the Bloch sphere. A general one-qubit Hamiltonian is described as
\begin{align}
    H_\text{1q}
 &= \sum_{\alpha=0}^{3} d_\alpha \hat{\sigma}^\alpha,
 \label{eq:Ham_1q}
\end{align}
where $d_\alpha \in \mathbb{R}$ for a Hermitian Hamiltonian, and the Pauli matrices are denoted by $\hat{\sigma}^0 = \mathds{1} = \smqty(\pmat{0})$, $\hat{\sigma}^1 = X = \smqty(\pmat{1})$, $\hat{\sigma}^2 = Y = \smqty(\pmat{2})$, and $\hat{\sigma}^3 = Z = \smqty(\pmat{3})$. For simplicity, consider $d_0 = d_3 = 0$ and denote the complex number $d_1 + id_2 = d e^{i\theta} = d\cos\theta + id\sin\theta$, where $d = \sqrt{d_1^2 + d_2^2}$ and $\theta = \Arg(d_1 + id_2)\in (-\pi, \pi]$. 
Then,
\begin{align}
    H_\text{1q}
 &= d_1X + d_2Y
  = d(\cos\theta X + \sin\theta Y).
\end{align}
The exact dynamics is given by
\begin{align}
     U(t) 
  &= e^{-itH_\text{1q}} \notag\\
  &= \cos(td)\mathds{1} - i\sin(td)(d_1 X + d_2 Y)/d.
\end{align}
In the last step, we used the SU(2) formula for the exponential of a Pauli vector 
\begin{align}
     e^{-i\phi \vu*{n} \vdot \vb*{\sigma}} 
  &= \mathds{1}\cos\phi - i(\vu*{n} \vdot \vb*{\sigma})\sin\phi.
  \label{eq:SU2Exp}
\end{align}

The second-order linear combination of Trotter unitaries (LCTU) is given by the Jordan--Trotter product defined in \cref{sec:generator} as follows:
\begin{subequations}
\begin{align}
  &   U_+(t) 
   = \tfrac{1}{2}\qty( e^{-itd_1 X} e^{-itd_2 Y} + e^{-itd_2 Y} e^{-itd_1 X} )
   ={} \\ & \tfrac{1}{2}\qty{ \cos[t(d_1+d_2)] \mathds{1} -i (X+Y) \sin[t(d_1+d_2)] } +{} \notag\\
     & \tfrac{1}{2}\qty{ \cos[t(d_1-d_2)] \mathds{1} -i (X-Y) \sin[t(d_1-d_2)] }. 
\end{align}
\end{subequations}

To characterize the approximation error, we compute the Frobenius norm\footnote{The Frobenius norm is defined as $\norm{A}_\text{F} = [\Tr(A^\dagger A)]^{1/2}$. Note that the operator norm is an upper bound of the Trotter error for state time evolution $\norm{\delta U \ket{\Psi}} \leq \norm{\delta U} \norm{\ket{\Psi}} = \norm{\delta U}$, assuming $\ket{\Psi}$ is normalized.}
as follows:
\begin{align}
  &\mathrel{\phantom{=}} \norm{U_+(t) - U(t)} \notag \\
  &= \begin{cases}
       0, 
         &\theta \in \qty{0, \pi, \pm\frac{\pi}{2}}\\
       \frac{\sqrt{2}}{6}\abs{(td)^3\sin(2\theta)} + \order{t^5},
         &\theta \notin \qty{0 , \pi, \pm\frac{\pi}{2}}.
     \end{cases}
\end{align}

On the other hand, the conventional second order Trotter $U_{(2)}(t)$ is given by
\begin{subequations}
\begin{align}
  &
     U_{(2)}(t) 
   = e^{-itd_1 X/2} e^{-itd_2 Y} e^{-itd_1 X/2}
   = {}\\ & \cos(td_1)\cos(td_2)\mathds{1} - i\cos(td_2)\sin(td_1)X - i\sin(td_2)Y.
\end{align}
\end{subequations}

Comparing to the SU(2) formula~\cref{eq:SU2Exp}, we find the corresponding generator discussed in \cref{sec:generator} as follows:
\begin{subequations}
\begin{align}
     z_{(2)} 
  &\equiv \log(U_{(2)})
   = -i\phi_{(2)}\qty(n^x_{(2)} X + n^y_{(2)} Y), \\
     \cos\phi_{(2)}
  &= \cos(td_1)\cos(td_2), \\
     n^x_{(2)}
  &= \cos(td_2)\sin(td_1)/\sin\phi_{(2)}, \\
     n^y_{(2)}
  &= \sin(td_2)/\sin\phi_{(2)}.
\end{align}
\end{subequations}

The nonunitary $U_+(t)$ admits a polar decomposition $U_+ = e^{-\alpha} U'_+$, where $U'_+$ is unitary, and the generator for $U_+$ is
\begin{subequations}
\begin{align}
     z_+
  &\equiv \log(U_+)
   = -i\phi_+\qty(n^x_+ X + n^y_+ Y) - \alpha\mathds{1}, \\
     \alpha
  &= -\frac{1}{2} \log\qty[1 - \sin^2(td_1)\sin^2(td_2)] > 0,\\
     \cos\phi_+
  &= e^{-\alpha}\cos(td_1)\cos(td_2) \in (-1, 1), \\
     n^x_+
  &= e^{-\alpha}\cos(td_2)\sin(td_1)/\sin\phi_+, \\
     n^y_+
  &= e^{-\alpha}\cos(td_1)\sin(td_2)/\sin\phi_+.
\end{align}
\end{subequations}
If $\sin^2(td_1)\sin^2(td_2) \to 1^+$, $\alpha \to +\infty$ and $U_+ \to 0$, which indicates large Trotter error. However, this will not happen if $|td_1|, |td_2| \ll 1$. The success probability is $P_{0\to 0} = \norm{U_+\ket{\Psi}}^2 = \ev{U_+^\dagger U_+}{\Psi} = e^{-2\alpha} = 1 - \sin^2(td_1)\sin^2(td_2) = 1 - d_1^2d_2^2t^4 + \order{t^6}$, which is independent of the state vector $\ket{\Psi}$. \Cref{eq:P00_t_expansion} gives $P_{0\to 0} \geq 1 - d_1^2d_2^2t^4/{2}$ (because $\norm{[A,B]}^2 = d_1^2 d_2^2 \norm{Z}^2 = 2d_1^2 d_2^2$). If successful, the normalized state after measurement of ancilla is $\ket{\Psi(t)} = U_+(t)\ket{\Psi}/\sqrt{P_{0\to 0}} = U'_+(t) \ket{\Psi}$, so $U'_+(t)$ is the effective unitary of the LCTU algorithm.

Finally, the norm distance between $U_+$ and the unitary $U'_+$ is $\norm{U_+(t) - U'_+(t)} = \frac{\sqrt{2}}{8}(td)^4 \sin^2(2\theta)$, which is not to the same order of the squared Trotter error $\norm{U_+(t) - U(t)}$ as it would be for the multiproduct formulas (see the top of p.~5 of Ref.~\onlinecite{Childs2012}). 

The error of the second-order Trotter is given by the operator norm
\begin{align}
  &\mathrel{\phantom{=}} \norm{U_{(2)}(t) - U(t)} \notag \\
  &= \begin{cases}
       0, 
         &\theta \in \qty{0, \pi, \pm\frac{\pi}{2}};\\
       \frac{\sqrt{5-3\cos(2\theta)}}{12}\abs{(td)^3\sin(2\theta)} + \order{t^5},
         &\theta \notin \qty{0 , \pi, \pm\frac{\pi}{2}}.
     \end{cases}
\end{align}

The errors are plotted in \cref{fig:1qTrotterError}. Since $\sqrt{2}/12 = \sqrt{5-3}/12 \leq \sqrt{5-3\cos(2\theta)}/12 \leq \sqrt{5+3}/12 = \sqrt{2}/6$, it follows that the error of second-order Trotter is always smaller than the two-unitary LCTU for this model. However, if $\theta = \frac{\pi}{4}$, the $X \leftrightarrow Y$ symmetry of the $H_\text{1q}$ conserved by the $U_+(t)$ evolution can have certain advantage over $U_{(2)}$ that breaks this symmetry. For instance, when acted on the eigenstates of $H_\text{1q}$, the second-order LCTU $U_+(t)$ gives a correct dynamic phase with $t^3$ error and keeps the eigenstates stationary on Bloch sphere as the exact evolution $U(t)$ does, but $U_{(2)}(t)$ will rotate these states on Bloch sphere.

\subsection{Symmetrized Hamiltonian Variational Ansatz for the Heisenberg Chain} \label{sec:apps-symHVA}
Solving the ground state (GS) and the ground state energy (GSE) of Hamiltonians of general quantum systems is a challenging problem and typically Quantum Merlin--Arthur (QMA) hard \cite{Watrous2008, Schuch2009}, which is quantum analogous to NP hard in classical computing. The variational quantum eigensolver (VQE) is the most practical quantum-classical hybrid algorithm to tackle this problem on NISQ quantum hardware. Among a plethora of ansatzes implementing VQEs, the Hamiltonian variational ansatz (HVA) stands out as a physics informed approach. It utilizes a single unitary product composed of factors derived from a one-step Trotter formula, enabling the time-evolution unitary of a Hamiltonian. The HVA has demonstrated significant strength in solving the GS problems from chemistry and physics~\cite{Wecker2015} as it encodes adiabatic evolutions for these general systems as the quantum approximate optimization algorithm (QAOA) does for the Ising spin models.

To improve algorithmic efficiency and reduce both quantum and classical computing resources, symmetry-preserving (-adapted) VQEs have been recently proposed\cite{Seki2020}. However, existing implementations often incur large circuit depth and gate budget overhead for symmetry preservation or projection. Inspired by the commutative symmetry of Jordan algebras, we propose a symmetrized HVA, termed symHVA using LCTU products. Applying symHVA to the 8-site Heisenberg spin chain, we find significantly improved performance compared to HVA. 

First, we define the Heisenberg model before showing the symHVA result compared with conventional HVA. The general spin-$\frac{1}{2}$ Heisenberg XYZ model [without local Zeeman ($h_l^z S_l^z$ terms) or transverse ($h_l^x S_l^x$ and $h_l^y S_l^y$ terms) fields] on an $L$-site chain is defined as
\begin{align}
     H_\text{XYZ}
  &= \sum_{l=0}^{L-1} \qty(J_x S_l^x S_{l+1}^x + J_y S_l^y S_{l+1}^y + J_z S_l^z S_{l+1}^z),
  \label{eq:xyzHeisenberg}
\end{align}
where $S_l^a = \frac{\hbar}{2}\sigma_l^a$, for $a\in \{x,y,z\}$. 
The periodic boundary condition (i.e., a circular chain) is considered in \cref{eq:xyzHeisenberg}, so $l+1 \equiv 0 \pmod{L}$ for $l=L-1$ in the sum (for open boundary condition, the sum upper limit is $l=L-2$). From now on, we consider the special case with isotropic antiferromagnetic interactions $J_x = J_y = J_z \equiv J > 0$ (i.e., the XXX Heisenberg model) and choose the units of energy so that $\frac{J\hbar}{4} = 1$. Then, the simplified antiferromagnetic Heisenberg model (AFHM) is
\begin{align}
     H
  &= \sum_{l=0}^{L-1} (X_l X_{l+1} + Y_l Y_{l+1} + Z_l Z_{l+1}) \label{eq:AFHM}\\
  &\equiv \sum_{l=0}^{L-1} H_l 
   = \sum_{l=0}^{L-1} \vb*{\sigma}_l \cdot \vb*{\sigma}_{l+1} \\
  &= \sum_{\text{even $l$}} \vb*{\sigma}_l \cdot \vb*{\sigma}_{l+1}
    +\sum_{\text{odd  $l$}} \vb*{\sigma}_l \cdot \vb*{\sigma}_{l+1}
   \equiv H_A + H_B . \label{eq:AFHM_A+B}
\end{align}
In the above, $\vb*{\sigma}_l \equiv (X_l, Y_l, Z_l)$. For an even number of sites $L$, the chain is a bipartite lattice, and the $A$ ($B$) part defined in \cref{eq:AFHM_A+B} is referred as the $A$ ($B$) sublattice.

\begin{figure}
  \centering
  \includegraphics[width=0.95\columnwidth]{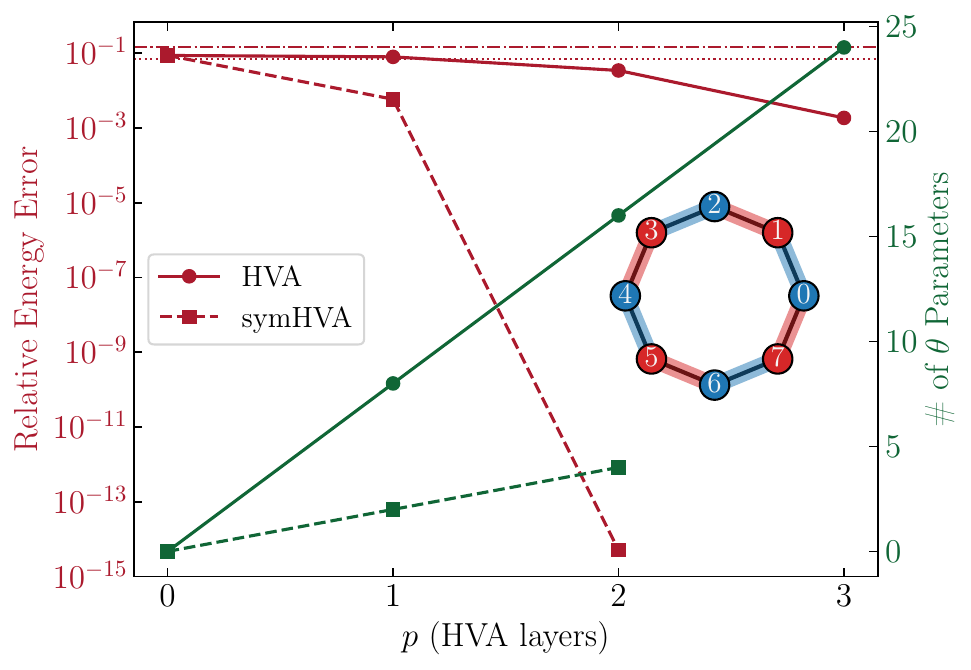}
\caption{Red lines, left axis: relative error between optimized GSE and the exact  one as a function of the number of HVA layers. Green lines, right axis: total number of variational parameters used in HVA and symHVA. Inset: graph representation of the 8-site AFHM. Two red horizontal lines are two reference energies relative to the GSE: the first excited energy (dashed-dotted line) and the energy at the middle point of the gap (dotted line).}
  \label{fig:8siteAFHM}
\end{figure}

We solve our AFHM ground state problem by 
\begin{align}
  \min_{\vb*{\theta}} \frac{ \ev{U^\dagger(\vb*{\theta}) \hat{H} U(\vb*{\theta})}{\psi_0} }
  { \ev{U^\dagger(\vb*{\theta}) U(\vb*{\theta})}{\psi_0} }.
\end{align}
We choose a good initial trial state $\ket{\psi_0} = \frac{2}{3}({\ket{\psi_{0A}} + \ket{\psi_{0B}}})$, where $\ket{\psi_{0A}}$ ($\ket{\psi_{0B}}$) is a valence bond state with singlet bonds on blue (red) edges as shown by an 8-vertex cycle graph inset in \cref{fig:8siteAFHM}. [The $\frac{2}{3}$ normalization factor is due to that the two valence bond states $\ket{\psi_{0A}}$ and $\ket{\psi_{0B}}$ are not orthogonal.] $U(\vb*{\theta})$ is unitary in HVA, with each circuit layer given by
  ${\exp}(i\sum_l \theta_l^p H_l)$
(total $8p$ parameters for $p$ layers); in symHVA, $U(\vb*{\theta})$ is nonunitary with each layer given by
\begin{align}
     U_{A,B}(\theta_1^p, \theta_2^p) 
  &=\frac{1}{2} \qty( e^{i\theta_1^p H_A} e^{i\theta_2^p H_B} + 
           e^{i\theta_1^p H_B} e^{i\theta_2^p H_A})    
\end{align}
with total $2p$ parameters. $p=2$ is sufficient for symHVA to reach the exact ground state energy (GSE) within (classical) machine precision, as shown in \cref{fig:8siteAFHM}.

\subsection{Mermin Polynomial Measurement for Multipartite Entanglement} \label{sec:apps-MerminPoly}

Above, we saw how symmetries can impact dynamical processes in three different settings. We turn our attention to symmetric quantum entangled measurements. By again combining symmetries and the encoding gadget of Fig.~\ref{fig:circuit}, this subsection demonstrates a duality between destructive and nondestructive ancilla-mediated measurements. We then examine how the latter gives rise to an efficient quantum circuits for Mermin polynomial evaluation.

Mermin polynomials appear in the Mermin-Klyshko (MK) inequalities~\cite{Mermin1990b} that are a multipartite generalization of the two-particle Clauser-Horne-Shimony-Holt (CHSH) inequality. Mermin~\cite{Mermin1990b} (see \cref{eq:MerminPolyA} below) originally proposed the degree-$n$ homogeneous polynomial
\begin{subequations}
\label{eq:MerminPolyA}
\begin{align}
     A_n
  &= \frac{1}{2i} \qty[
     \prod_{l=1}^{n}\qty(\sigma_l^x +i\sigma_l^y) -
     \prod_{l=1}^{n}\qty(\sigma_l^x -i\sigma_l^y)] \label{eq:MerminPolyA1} \\
  &= \sum_{(\vb*{p}^x, \vb*{p}^y)\in \mathcal{I}} (-1)^{\frac{|\vb*{p}^y|_1-1}{2}}
  \prod_{l=1}^{n} (\sigma_l^x)^{p_l^x}(\sigma_l^y)^{p_l^y}, \label{eq:MerminPolyA2}
\end{align}
\end{subequations}
where the index set $\mathcal{I} = \{(\vb*{p}^x, \vb*{p}^y) \mid \ (p_l^x,p_l^y)=(1,0) \text{ or } (0,1)  \text{ and } |\vb*{p}^y|_1 \text{ is odd}\}$ with $\vb*{p}^a \equiv (p_1^a, \dots, p_n^a)$ for $a=x,y$ and $|\vb*{p}^y|_1 \equiv \sum_{l} |p_l^y|$. Here, $(\sigma_l^x, \sigma_l^y)$ are $n$ pairs of local spin operators. \Cref{eq:MerminPolyA2} directly follows \cref{eq:MerminPolyA1} by binomial expansion. The Mermin polynomial $A_n$ includes $|\mathcal{I}| = 2^{n-1}$ terms of degree-$n$ monomials, i.e., Pauli strings consisting of either $\sigma^x$ or $\sigma^y$ for each qubit. Because only odd number of $\sigma^y$'s appear, these Pauli strings mutually commute and hence $A_n$ is a stabilizer Hamiltonian. Because each term of the stabilizer Hamiltonian $A_n$ has $\pm 1$ coefficient, $\ev{A_n}{\Psi} \leq \norm{A_n} \leq \mathcal{I} = 2^{n-1}$. The maximal quantum mechanics value $\ev{A_n}_\text{max} = 2^{n-1}$ is achieved with Greenberger--Horne--Zeilinger (GHZ) state $\ket{\Psi_+} = \frac{1}{\sqrt{2}}(\ket{\vb{0}} + i\ket{\vb{1}})$~\cite{Mermin1990b} ($\ket{\vb{0}} \equiv \ket{00\cdots 0}$ and $\ket{\vb{1}} \equiv \ket{11\cdots 1}$), while, by symmetry, the minimal value $\ev{A_n}_\text{min} = -2^{n-1}$ is achieved with GHZ state $\ket{\Psi_-} = \frac{1}{\sqrt{2}}(\ket{\vb{0}} - i\ket{\vb{1}})$. $\ket{\Psi_-}$ is a frustration-free ground state since it simultaneously minimizes each stabilizer term and $\ket{\Psi_+}$ can be viewed as frustration-free ground state for $-A_n$.

It is also interesting to note that $A_n^4 = 2^{2n-2} A_n^2$ by simple calculation, and thus, in addition to the eigenvalues $\pm 2^{n-1}$, all other eigenvalues of $A_n$ is 0. We find numerically this null space has $2^n - 2$ dimensions and the remaining two dimensions of the Hilbert space correspond to $\{\ket{\Psi_\pm}\}$ with eigenvalues $\pm 2^{n-1}$. This indicates $A_n$ projects into GHZ state subspace when acting on any give state $\ket{\Psi}$, and the eigendecomposition $A_n = 2^{n-1}(\op{\Psi_+} - \op{\Psi_-}) = 2^{n-1}(i\op{\vb{1}}{\vb{0}} - i\op{\vb{0}}{\vb{1}})$. The last expression can be easily obtained by using $\sigma_l^x + i\sigma_l^y = 2\op{0}{1}_l$ and $\sigma_l^x - i\sigma_l^y = 2\op{1}{0}_l$ in \cref{eq:MerminPolyA1}.

Collins \textit{et al.}~\cite{Collins2002} defined the more general and most commonly used Mermin polynomials for general $n$ pairs of single-particle (-qubit) operators $\{(a_l, a'_l) \mid l=1,2,\ldots, n\}$ using the following recursive relations:
\begin{subequations}
  \label{eq:MerminP}
\begin{align}
  M_k  &= \frac{1}{2}M_{k-1}(a_k + a'_k) + \frac{1}{2}M'_{k-1}(a_k - a'_k), \label{eq:MerminP1}\\
  M'_k &= \frac{1}{2}M'_{k-1}(a'_k + a_k) + \frac{1}{2}M_{k-1}(a'_k - a_k), \label{eq:MerminP2}
\end{align}
\end{subequations}
where $M_1 = a_1$ and $M'_1 = a'_1$.
Note that $M'_k$ is obtained by exchanging all the primed and nonprimed $a$'s. 
Here, we define all $a$ operators as $n$-qubit tensor products $a_l = \mathds{1}\otimes \cdots \mathds{1} \otimes a \otimes \mathds{1} \cdots \otimes \mathds{1}$, where $a$ acts on the $l$-th qubit and similarly for primed $a$'s. Thus, we have $[a_k, a_l] = 0$, $[a'_k, a'_l] = 0$, for all indices and $[M_k, a_l]=0$ and $[M_k, a'_l]=0$ for $k < l$. Finally, $a$ and $a'$ are often assumed to be dichotomic (involution) operators: $a_k^2 = {a'_k}^2 = \mathds{1}$ for all $k$.

Before proceeding, we remark the difference between Collins \textit{et al.}'s $M_n$ and Mermin's $A_n$:
(1) If we assume the special setting $a_l = \sigma^x_l$ and $a'_l = \sigma^y_l$ in \cref{eq:MerminP_cf} we find $M_n^4 = 2^{n-1} M_n^2$, so the eigenvalues of $M_n$ are $0, \pm 2^{(n-1)/2}$. The maximal quantum mechanics value $\ev{M_n}_\text{max} = 2^{(n-1)/2}$, which is given by the largest eigenvalue and agrees with Collins \textit{et al.}~\cite{Collins2002}. 
(2) The classical maximal value by local realism $\ev{M_n}_\text{LR} = 1$~\cite{Collins2002,Alsina2016}, while $\ev{A_n}_\text{LR} = 2^{n/2}$ for even $n$ and $\ev{A_n}_\text{LR} = 2^{(n-1)/2}$ for odd $n$~\cite{Mermin1990b}. The classical value difference is due to different normalization: all $2^{n-1}$ commuting monomial terms in $A_n$ have coefficient $\pm 1$, while for $M_n$ the $2^{n}$ ($2^{n-1}$) monomial terms have coefficients $\pm 2^{-n/2}$ ($\pm 2^{-(n-1)/2}$) for even (odd) $n$. For even $n$, the $2^n$ terms in $M_n$ do not mutually commute, so the quantum mechanics maximal value $\ev{M_n}_\text{QM} = 2^{(n-1)/2} < 2^n\times 2^{-n/2}$, different from the frustration-free case where each term is optimized by the GHZ states.
(3) Similar to $A_n$, using \cref{eq:MerminP_cf1}, we find the eigendecomposition $M_n = (1+i)^{n-1}\op{\vb{1}}{\vb{0}} + (1-i)^{n-1}\op{\vb{0}}{\vb{1}} = 2^{(n-1)/2}(\op*{\tilde{\Psi}_+} - \op*{\tilde{\Psi}_-})$, where $\ket*{\tilde{\Psi}_\pm} = \frac{1}{\sqrt{2}} (\ket{\vb{0}} \pm e^{i\pi(n-1)/4} \ket{\vb{1}})$, the corresponding GHZ states for $M_n$. The eigendecompositions of $A_n$ and $M_n$ suggest that these operators can be measured with multi-qubit swap test~\cite{Garcia-Escartin}: $\ev{A_n}{\Psi} = 2^{n-1} (\abs*{\ip*{\Psi_+}{\Psi}}^2 - \abs*{\ip*{\Psi_-}{\Psi}}^2)$ and $\ev{M_n}{\Psi} = 2^{(n-1)/2} (\abs*{\ip*{\tilde{\Psi}_+}{\Psi}}^2 - \abs*{\ip*{\tilde{\Psi}_-}{\Psi}}^2)$, where each of the two overlaps can be measured with swap test between a GHZ state and an input state $\ket{\Psi}$. We will derive an efficient quantum circuit for measuring Mermin polynomials in general $(a_l, a'_l)$ settings without this simple eigendecomposition.

To proceed, we rewrite \cref{eq:MerminP} using the following matrix notation.
\begin{align}
  \pmqty{M_1 \\ M'_1}
  &= \pmqty{a_1 \\ a'_1}
   = \frac{1}{2} 
     \pmqty{a_1+a'_1  & a_1 - a'_1 \\ -a_1+a'_1  & a_1 + a'_1} 
     \pmqty{1 \\ 1}. \\  
  \pmqty{M_k \\ M'_k}
  &= \frac{1}{2} 
     \pmqty{a_k+a'_k  & a_k - a'_k \\ -a_k+a'_k  & a_k + a'_k} 
     \pmqty{M_{k-1} \\ M'_{k-1}} \label{eq:MerminP3}\\
  &= \frac{1}{\sqrt{2}}\qty( a_k e^{\frac{i\pi}{4}Y} + a'_k e^{\frac{-i\pi}{4}Y}) 
     \pmqty{M_{k-1} \\ M'_{k-1}} \\
  &= \cdots \notag \\
  &= 2^{-\frac{k}{2}}\prod_{l=1}^{k} \qty( a_l e^{\frac{i\pi}{4}Y} + a'_l e^{\frac{-i\pi}{4}Y}) \pmqty{1 \\ 1}, \label{eq:MerminP4}
\end{align}
where $Y = \smqty(\pmat{2})$ is the Pauli matrix acting on an ancilla qubit that encodes the algebraic recursive relation \footnote{Note that using the matrix form for $Y$ does \emph{not} mean we have chosen the computation basis for the system qubits}. For convenience, we refer to the matrix operator $\frac{1}{2}\smqty(a_k+a'_k  & a_k - a'_k \\ -a_k+a'_k  & a_k + a'_k) = \frac{1}{\sqrt{2}} (a_k e^{\frac{i\pi}{4}Y} + a'_k e^{\frac{-i\pi}{4}Y}) \equiv \mathcal{A}_k$ as the transfer matrix.

The algebraic symmetry, derived from the interchange of all primed and nonprimed $a$'s in \cref{eq:MerminP3}, becomes more transparent when expressed in terms of symmetric and antisymmetric combinations. This transparency is achieved by transforming $(M_k, M'_k)$ into $\frac{1}{2}(M_k+M'_k, -M_k+M'_k) \equiv (M_k^+, M_k^-)$ through the left multiplication of $\frac{1}{\sqrt{2}} e^{\frac{i\pi}{4}Y}$ on both sides of \cref{eq:MerminP3}.
Given that  $[\mathcal{A}_k, e^{\frac{i\pi}{4}Y}] = 0$, the transfer matrix stays the same: $\mathcal{A}_k = \frac{1}{\sqrt{2}} (a_k e^{\frac{i\pi}{4}Y} + a'_k e^{\frac{-i\pi}{4}Y}) = \mathds{1} a_k  ^+ - iY a_k^-$, where $a_k^\pm = (\pm a_k + a'_k)/2$. Then, the recursive formula becomes
\begin{align}
     \pmqty{M_1^+ \\ M_1^-} 
  &= \mathcal{A}_0 \pmqty{1 \\ 0}, \\
     \pmqty{M_k^+ \\ M_k^-}
  &= \mathcal{A}_k \pmqty{M_{k-1}^+ \\ M_{k-1}^-} \\
  &= \qty(\mathds{1} a_k^+ - iY a_k^-) 
     \pmqty{M_{k-1}^+ \\ M_{k-1}^-} \label{eq:MerminP+-} \\
  &= \cdots \notag\\
  &= \prod_{l=1}^{k} \qty(\mathds{1} a_l^+- iY a_l^-) \pmqty{1 \\ 0}.
  \label{eq:MerminP+-1}
\end{align}
The $\pm$ superscripts indicate the sign change under the symmetry of exchanging all the primed and nonprimed $a$'s, so $(M_l^+, M_l^-)^T \to (M_l^+, -M_l^-)^T = Z (M_l^+, M_l^-)^T$ and $\mathcal{A}_k = \mathds{1} a_k^+ - iY a_k^- \to \mathds{1} a_k^+ + iY a_k^- = Z \mathcal{A}_k Z$, where $Z = \smqty(\pmat{3})$. Thus, \cref{eq:MerminP+-} still holds under the symmetry of exchanging all the primed and nonprimed $a$'s as it should be.

First, we point out some similarity between \cref{eq:MerminP3} and the unitary operator $\mathcal{U}$ in the boxed part in the circuit in \cref{fig:circuit} that is given by
\begin{align*}
     \mathcal{U}
  &= H\otimes I \cdot \Bigl(\op{0}_a \otimes U + \op{1}_a \otimes V \Bigr) \cdot H\otimes I \\
  &= \op{+}_a \otimes U + \op{-}_a \otimes V \\
  &= \frac{1}{2}\pmqty{U+V & U-V  \\ U-V &U+V} \tagthis \label{eq:UV_LCU}.
\end{align*}
The operators $\op{\pm} = (\mathds{1}\pm X)/2$ are both projection operators. Consequently,  $\mathcal{U}$ is unitary, satisfying the condition $\mathcal{U}\mathcal{U}^\dagger = \op{+}_a \otimes (UU^\dagger) + \op{-}_a \otimes (VV^\dagger) = \mathds{1}_a \otimes \mathds{1}$. In \cref{eq:MerminP3}, the transfer matrix $\mathcal{A}_k$ is similar to \cref{eq:UV_LCU} except for the minus sign of the $(2,1)$ matrix element. Due to this minus sign, this matrix operator is not unitary, so we need a second ancilla qubit to be used with the LCU for this matrix operator. The first ancilla deterministically encodes the mutual recursions and the second ancilla probabilistically block-encodes the nonunitary into a unitary via LCU and measurements. However, we can solve the recursion described in \cref{eq:MerminP+-1} in a closed form using the $Y$ eigenbasis $\qty{\ket{+_y} = \frac{1}{\sqrt{2}}\spmqty{1 \\ i}, \ket{-_y} = \frac{1}{\sqrt{2}}\spmqty{1 \\ -i}}$ to preserve the first qubit. We multiply the following resolution of identity:
\begin{align*}
     \mathds{1}
  &= \frac{\mathds{1} + Y}{2} + \frac{\mathds{1} - Y}{2} \\
  &= \op{+_y} + \op{-_y}
   = \pmqty{\ket{+_y} & \ket{-_y}} \pmqty{\bra{+_y} \\ \bra{-_y}}
\end{align*}
at both ends of the product in \cref{eq:MerminP+-1}, and find
\begin{align*}
     \pmqty{M_k^+ \\ M_k^-}
  &= \pmqty{\ket{+_y} & \ket{-_y}} \pmqty{\bra{+_y} \\ \bra{-_y}} \\
  &\phantom{={}}\times   \prod_{l=1}^{k} \qty[\qty(\mathds{1} a_l^+ - iY a_l^-) 
     \pmqty{\ket{+_y} & \ket{-_y}} \pmqty{\bra{+_y} \\ \bra{-_y}}]
     \pmqty{1 \\ 0} \\
  &= \pmqty{\ket{+_y} & \ket{-_y}}  
     \pmqty{\prod_{l=1}^{k} (a_l^+ - i a_l^-) &0 \\
            0 &\prod_{l=1}^{k} (a_l^+ + i a_l^-)} \\ 
  &\phantom{={}}\times   \pmqty{\bra{+_y} \\ \bra{-_y}} \pmqty{1 \\ 0} \\
  &= \frac{1}{\sqrt{2}} \pmqty{\ket{+_y} & \ket{-_y}}  
     \pmqty{\prod_{l=1}^{k} (a_l^+ - i a_l^-) \\
            \prod_{l=1}^{k} (a_l^+ + i a_l^-)} \\
  &= \frac{1}{2} \pmqty{1 & 1 \\ i &-i }
     \pmqty{\prod_{l=1}^{k} (a_l^+ - i a_l^-) \\
            \prod_{l=1}^{k} (a_l^+ + i a_l^-)}.
\end{align*}
Therefore,
\begin{align*}
     \pmqty{M_k \\ M'_k}
  &= \pmqty{1 & -1 \\ 1 &1 } \pmqty{M_k^+ \\ M_k^-} \\
  &= \qty(\frac{1-i}{2})^{k+1} \pmqty{1 & i \\ i &1 }
     \pmqty{\prod_{l=1}^{k} (a'_l + i a_l) \\
            \prod_{l=1}^{k} (a_l + i a'_l)},
\end{align*}
where we have used $a_l^+ \pm i a_l^- = \frac{1\mp i}{2}(a_l \pm i a'_l)$. The symmetry of exchanging all the primed and nonprimed symbols is equivalent to swapping the two rows of the vectors $(M_k, M'_k)^T$ and $(\prod_{l=1}^{k} (a'_l + i a_l), \prod_{l=1}^{k} (a_l + i a'_l))^T$, which is the action of $X = \spmqty{\pmat{1}}$ operator. The recursive formula respects this symmetry because $\spmqty{1 &i \\ i &1}$ and $X$ operator commute.

Finally, we have the following closed-form expressions for all $M_k$ and $M'_k$ defined the by the recursion in \cref{eq:MerminP}
\begin{subequations}
  \label{eq:MerminP_cf} 
\begin{align}
     M_k 
  &= \frac{(1-i)^{k+1}}{2^{k+1}}\qty[i\prod_{l=1}^{k} (a_l + i a'_l) 
     + \prod_{l=1}^{k} (a'_l + i a_l)], \label{eq:MerminP_cf1} \\
     M'_k 
  &= \frac{(1-i)^{k+1}}{2^{k+1}}\qty[ \prod_{l=1}^{k} (a_l + i a'_l) 
     +i\prod_{l=1}^{k} (a'_l + i a_l)]. \label{eq:MerminP_cf2}
\end{align}
\end{subequations}
With \cref{eq:MerminP_cf}, we can easily write down the closed-form for the Svetlichny polynomial $S_k$ defined in Ref.~\onlinecite{Collins2002} [Eq.~(13) therein]. For $k$ odd, $S_k = \frac{1}{2} (M_k + M'_k)$ takes an especially symmetric form as
\begin{align}
\label{eq:symm-form}
     S_k 
  &= \frac{(1-i)^{k}}{2^{k+1}}\qty[
       \prod_{l=1}^{k} (a_l + i a'_l) 
     + \prod_{l=1}^{k} (a'_l + i a_l)].
\end{align}

We can check \cref{eq:MerminP_cf1} for $k=3$ as follows:
\begin{align*}
     M_3 
  &= \frac{(1-i)^4}{2^4}
     \left[i(a_1 + ia'_1)(a_2 + ia'_2)(a_3 + ia'_3)\right. \\
  &\mathrel{\phantom{=}}\left. +i^3 (a_1 - ia'_1)(a_2 - ia'_2)(a_3 - ia'_3) \right] \\
  &= \frac{1}{2}(a'_1 a_2 a_3 + a_1 a'_2 a_3 + a_1 a_2 a'_3 - a'_1 a'_2 a'_3) .
\end{align*}
This result agrees with Eq.~(5) in Ref.~\onlinecite{Collins2002}.

\subsubsection{Mermin Polynomial Measurement Circuits}
For an $n$-qubit quantum state $\ket{\Psi}$, to measure $\mel{\Psi}{M_n}{\Psi}$ with the closed form of \cref{eq:MerminP_cf1}, we can first measure the two product terms separately. \Cref{eq:merminPolyM_cf1_circ} shows how to apply the $n$ factors $(a_l + ia'_l)/2$ ($l=1, \dots, n$) in the first product term to $\ket{\Psi}$ (similar for the second term).
\begin{align}
 \mspace{18mu}
 \begin{aligned} \small
  \Qcircuit @C=0.2em @R=0.8em {%
   \lstick{\ket{0}_a\mspace{-5mu}}
   &\gate{H} &\ctrlo{1} &\ctrl{1} &\gate{H} &\meter &\cds{3}{\text{\large \bf $\ddots$}}
   &\gate{H} &\ctrlo{3} &\ctrl{3} &\gate{H} &\meter \\
   &\lstick{\ket{q_1}\mspace{-5mu}} &\gate{a^{\phantom{\prime}}_1} &\gate{ia'_1} &\qw &\ustick{ m_1 = 0} 
   & & & & & &\ustick{m_n = 0} \\
   & & & & & &
   & & & &\rstick{\;\ket{\Phi}} & &\\\
   & & & & & &
   &\lstick{\ket{q_n}\mspace{-5mu}} &\gate{a^{\phantom{\prime}}_n} &\gate{ia'_n} &\qw
   \gategroup{2}{11}{4}{11}{0.6em}{\}}
  }
 \end{aligned}
 \label{eq:merminPolyM_cf1_circ}
\end{align}
We recursively implement each factor with LCU for the sum of two terms $a_l$ and $i a'_l$. Thus, we only need one ancilla if we reuse and reset it in the iterative algorithm (in this specific case, reset is not necessary since the same LCU coefficients appear in all factors). The final \emph{normalized} $n$-qubit state in \cref{eq:merminPolyM_cf1_circ} is $\ket{\Phi} = p_{\vb{0}}^{-1/2}\prod_l [(a_l + ia'_l)/2]\ket{\Psi}$, where $p_{\vb{0}} = 2^{-n}\ev*{\prod_{l}(1 + \frac{i}{2}[a_l, a'_l])}{\Psi}$ is the probability of measuring all $0$s in the ancilla qubit ($p_{\vb{0}}$ obtained using the fact $\ip{\Psi} = 1$ and $a_l^2 = {a'_l}^2 = 1$). A swap test can determine $|{\ip{\Psi}{\Phi}}|^2$ and in principal gives the contribution of the first product term in \cref{eq:MerminP_cf1}. However, the phase of the overlap ${\ip{\Psi}{\Phi}}$ also needs to be accounted for in order to have the correct sum of the two terms. Since $M_n$ is Hermitian, we only need to evaluate $\Re[e^{-i\frac{\pi}{4}(k-1)}\ip{\Psi}{\Phi}]$ instead of running the circuit in \cref{eq:merminPolyM_cf1_circ} again to measurement the second product term.

To circumvent the difficulty of dealing with the complex phase of $\ip{\Psi}{\Phi}$, we now compile a circuit to apply both terms in $M_k$ at once. Using \cref{eq:MerminP4}, we find
\begin{align}
     M_n
  &= 2^{-\frac{n}{2}} \bra{0}_b
     \qty[\prod_{l=1}^{n} \qty (a_l e^{\frac{i\pi Y_b}{4}} + a'_l e^{\frac{-i\pi Y_b}{4}})]
     (\ket{0}_b + \ket{1}_b) \notag \\
  &= 2^{-\frac{n-1}{2}} \bra{0}_b
     \qty[\prod_{l=1}^{n} \qty (a_l - ia'_l Y_b)]
     e^{\frac{i(n-1)\pi Y_b}{4}}
     \ket{0}_b \notag \\
  &= 2^{\frac{n+1}{2}} \bra{0} e^{\frac{-i\pi X_b}{4}}
     \qty[\prod_{l=1}^{n} \frac{1}{2} \qty(a_l + ia'_l Z_b)] \notag \\
  &\mathrel{\phantom{=}}
     \times e^{\frac{-i(n-1)\pi Z_b}{4}}
     e^{\frac{i\pi X_b}{4}} \ket{0}_b,
  \label{eq:merminPolyFactorAncilla}
\end{align}
where $\ket{0}_b$ and $\ket{1}_b$ are computational basis states for the ancilla qubit (denoted with subscript $b$) that encodes the double recursive relations (for $M_n$ and $M'_n$), and the Pauli operators ($X_b$, $Y_b$, and $Z_b$) act on this qubit. In the above, we first used $e^{-i\pi Y/4} \ket{0} = (\ket{0} + \ket{1})/\sqrt{2}$ (equivalent to the action of an Hadamard gate), and then, $e^{i\pi X/4} Y e^{-i\pi X/4} = -Z$ to conjugate all $Y$ gate to $Z$ gate that most quantum hardware use. Since $e^{-i\pi X/4} \equiv e^{-i\pi/4} \sqrt{X}$, where $\sqrt{X}$ is a native gate for quantum hardware, we will replace $e^{-i\pi X/4}$ with $\sqrt{X}$ gate.

Finally, similar to the compilation of the circuit in \cref{eq:merminPolyM_cf1_circ}, we compile the operator $\bar{M}_n \equiv 2^{-(n+1)/2}M_n$ in \cref{eq:merminPolyFactorAncilla} into the following circuit by adding a $b$-ancilla qubit at the top to apply the simultaneously controlled gates $\op{0}_a \otimes \mathds{1} + \op{1}_a \otimes (i a'_l Z_b)$.
\begin{align}
 \mspace{18mu}
 \begin{aligned} \small
  \Qcircuit @C=0.2em @R=0.8em {%
   \lstick{\ket{0}_b\mspace{-5mu}}
   &\gate{\sqrt{X}^\dagger} &\gate{R_z} &\gate{Z}\qwx[1] &\qw &\qw &\qw 
   &\qw &\qw &\gate{Z}\qwx[1] &\gate{\sqrt{X}}  &\meter &\dstick{\mspace{-58mu} \raisebox{-0.9em}{$m_{n+1} = 0$}}  \\
   \lstick{\ket{0}_a\mspace{-5mu}}
   &\gate{H} &\ctrlo{1} &\ctrl{1} &\gate{H} &\meter &\cds{3}{\text{\large \bf $\ddots$}}
   &\gate{H} &\ctrlo{3} &\ctrl{3} &\gate{H} &\meter \\
   &\lstick{\ket{q_1}\mspace{-5mu}} &\gate{a^{\phantom{\prime}}_1} &\gate{ia'_1} &\qw &\ustick{ m_1 = 0} 
   & & & & & &\ustick{m_n = 0} \\
   & & & & & &
   & & & &\rstick{\;\ket{\Phi}} & &\\\
   & & & & & &
   &\lstick{\ket{q_n}\mspace{-5mu}} &\gate{a^{\phantom{\prime}}_n} &\gate{ia'_n} &\qw
   \gategroup{3}{11}{5}{11}{0.6em}{\}}
  }
 \end{aligned}
 \label{eq:merminPolyFactorCirc}
\end{align}

In \cref{eq:merminPolyFactorCirc}, the $R_z$ gate acting on the $b$-ancilla qubit is $R_z((n-1)\pi/2) = e^{-i(n-1)\pi Z_b/4}$. Now, the normalized state $\ket{\Phi} = p_{\vb{0}}^{-1/2} 2^{-(n+1)/2}   M_n \ket{\Psi}$ and $p_{\vb{0}} = 2^{-(n+1)} \ev*{M_n^2}{\Psi} \leq \frac{1}{4}$ [obtained using the fact $M_n^2 = 2^{n-1} (\op*{\tilde{\Psi}_+} + \op*{\tilde{\Psi}_-})$]. To measure the magnitude of the Mermin polynomial $\ev*{M_n}{\Psi} \in [-2^{(n-1)/2}, 2^{(n-1)/2}]$, we only need one swap-test $\order{n}$-depth circuit to estimate $|{\ip{\Psi}{\Phi}}|^2$ after applying $\order{n}$-depth circuit in \cref{eq:merminPolyFactorCirc} to obtain $\ket{\Phi}$. Thus, we can directly measure Mermin polynomial with only $\order{n}$-depth circuits and the success probability $p_{\vb{0}} = \order{1}$ for obtaining the state $\ket{\Phi}$ for high-fidelity states with respect to $\ket*{\tilde{\Psi}_\pm}$. In fact, $p_{\vb{0}} = \frac{1}{4}$ for any state in the two-dimensional subspace spanned by $\ket*{\tilde{\Psi}_\pm}$. Finally, because $\ev*{M_n}{\Psi}^2 \leq \ev*{M_n^2}{\Psi} = 2^{n+1}p_{\vb{0}}$, with \cref{eq:merminPolyFactorCirc} alone, we can efficiently measure an upper bound of Mermin polynomial $2^{(n-1)/2}\sqrt{4 p_{\vb{0}}} \geq |{\ev*{M_n}{\Psi}}|$.

\subsubsection{Mermin Polynomial Measurement Equivalence}
In the setting~\cite{Alsina2016a} $a_l=\sigma_l^x$ and $a'_l=\sigma_l^y$ for all qubits, the circuit in \cref{eq:merminPolyM_cf1_circ} [as well as \cref{eq:merminPolyFactorCirc}] can be simplified considerably. The factors appearing in $M_k$ takes the form of the ladder operators $a_l \pm ia'_l = \sigma_l^x \pm i\sigma_l^y=\sigma^\pm_l$, which is exactly the same factor appeared in the polynomial defined by Mermin~\cite{Mermin1990b} [see \cref{eq:MerminPolyA1}]. With a slight change of notation, we write $\sigma_l^x  \pm i\sigma_l^y \equiv X_l \pm iY_l = \frac{\mathds{1} \pm Z_l}{2}(2X_l)$, where the non-unitary part 
\begin{equation}
    U_\pm = \frac{\mathds{1} \pm Z_l}{2}
    \label{eq:projectionLCU}
\end{equation}
can be implemented using the circuit in \cref{fig:circuit} with $U=\mathds{1}/2$ and $V=Z_l/2$. Considering the first term of $U_\pm$ is the identity operator $\mathds{1}$, we simplify the circuit for $U_\pm X_l$ as follows. Using the identities
\begin{subequations}
\begin{align}
\begin{aligned} \small
  \Qcircuit @C=0.4em @R=0.4em {%
   &\gate{H} &\ctrl{1} &\gate{H} &\qw \\
   &\qw &\gate{Z} &\qw &\qw}
\end{aligned}
&= 
\begin{aligned} \small
  \Qcircuit @C=0.4em @R=1.2em {%
   &\targ &\qw\\
   &\ctrl{-1} &\qw}
\end{aligned} \label{eq:ZtoX}\\
\begin{aligned} \small
  \phantom{\ket{\psi}}
  \Qcircuit @C=0.4em @R=0.6em {%
   \lstick{\ket{0}}
   &\targ &\meter\\
   \lstick{\ket{\psi}}
   &\ctrl{-1} &\qw}
\end{aligned}
&=\; 
\begin{aligned} \small
  \phantom{\ket{\psi}}
  \Qcircuit @C=0.4em @R=0.4em {%
   \lstick{\ket{0}}
   &\targ &\qw\\
   \lstick{\ket{\psi}}
   &\ctrl{-1} &\meter}
\end{aligned}
 =\; 
\begin{aligned} \small
  \phantom{\ket{\psi}}
  \Qcircuit @C=0.4em @R=0.2em {%
   \lstick{\ket{0}}
   &\targ &\meter\\
   \lstick{\ket{\psi}}
   &\ctrl{-1} &\meter}
\end{aligned} \label{eq:measureBellState}\\
\begin{aligned} \small
  \Qcircuit @C=0.4em @R=0.4em {%
   &\gate{U} &\qw\\
   &\ctrl{-1} &\meter}
\end{aligned}
&=
\begin{aligned} \small
  \Qcircuit @C=0.4em @R=0.4em {%
   &\gate{U} &\qw\\
   &\meter \cwx[-1] &\cw}
\end{aligned}
 =
\begin{aligned} \small
  \Qcircuit @C=0.4em @R=0.4em {%
   &\qw  &\gate{U} &\qw\\
   &\meter &\control \cw \cwx[-1] &\cw}
\end{aligned} \label{eq:deferredMeasurement}
\end{align}
\end{subequations}
we find
\begin{align}
 &\begin{aligned} \small
  \Qcircuit @C=0.4em @R=0.4em {%
   \lstick{\ket{0}_a}
   &\gate{H} &\ctrl{1} &\gate{H} &\meter \\
   \lstick{\ket{q_l}}
   &\gate{X} &\gate{Z} &\qw      &\qw}
 \end{aligned} \label{eq:ancillaProjMeasCirc} \\
 \overset{(\ref{eq:ZtoX})}{=}\;
 &\begin{aligned} \small
  \Qcircuit @C=0.4em @R=0.4em {%
   &\qw      &\targ     &\meter \\
   &\gate{X} &\ctrl{-1} &\qw}
 \end{aligned}
 \overset{(\ref{eq:measureBellState})}{=}\; \begin{aligned} \small
  \Qcircuit @C=0.4em @R=0.4em {%
   &\qw      &\targ     &\qw \\
   &\gate{X} &\ctrl{-1} &\meter}
 \end{aligned} 
 \overset{(\ref{eq:deferredMeasurement})}{=}
 \begin{aligned} \small
  \Qcircuit @C=0.4em @R=0.4em {%
   &\qw      &\targ \\
   &\gate{X} &\meter \cwx[-1] }
 \end{aligned} \notag \\
 \sim\mspace{15mu}
 &\begin{aligned} \small
  \Qcircuit @C=0.4em @R=0.8em {%
   &\qw      &\rstick{\text{discard}} \qwa \\
   &\gate{X} &\meter }
 \end{aligned} 
\label{eq:projMeasCirc}
\end{align}
Among the identities used in the above, \cref{eq:measureBellState} is due to the equivalent disentangling effect of measuring any qubit of Bell- and GHZ-like states, and \cref{eq:deferredMeasurement} is due to the principal of deferred measurement. In the final step, the ancilla can be discarded since we do not need the information in the ancilla registry, and thus, we do not need the ancilla qubit from the beginning if we choose to perform measurement on the principal system.

The projective measurement on ancilla \cref{eq:ancillaProjMeasCirc} can also be re-interpreted as measurement on the qubit of main system as shown in \cref{eq:projMeasCirc}.
Hence, addressing the Mermin polynomial with Fig.~\ref{fig:circuit} naturally reveals the equivalence between direct measurement, on the principal system, and indirect measurement, on an environmental pointer state represented by ancilla qubits above.
The two theoretical equivalent options (measure a single ancilla iteratively vs measure all qubits in the main entangled system) can be exploited to optimize readout error, post-selection cost, and maybe even avoid certain measurement-related loopholes in quantum entanglement verification.


In the $a_l=\sigma_l^x$ and $a'_l=\sigma_l^y$ setting, from the circuit in \cref{eq:merminPolyFactorCirc} we see that, modulo a global bit-flip on the principal system, measurements in the computational basis provide a way to stochastically evaluate the Mermin polynomial. To make this concrete, we can use the circuit above to measure the $m_{i} = 0, 1$ value, corresponding to $\pm$ in \cref{eq:projectionLCU}, with \cref{fig:circuit} manifesting as an ancilla-assisted measurement circuit. The ancilla qubit measurement result reveals which bit string projection occurred on the principal system. In terms of the measurement outcome, $m_{i}$, dependent gate $U^{(i)}_{m_{i}}$ acting on the $i$th qubit, the Mermin polynomial, $A = 2^{N-1} (\otimes^N_{i=1} U^{(i)}_{0} X_i - \otimes_i U^{(i)}_{1} X_i)/i$, is expressed as the sum of two circuit outcomes from \cref{eq:merminPolyFactorCirc}. By instead factorizing an $iY_l$, such that $X_l \pm i Y_l = (Z \pm 1) iY_l$, this becomes $A = 2^{N-1} (\otimes^N_{i=1} U^{(i)}_{0} + \otimes_i U^{(i)}_{1}) \otimes_i Y_i$.


As a final brief note, recall the polynomial factorized as a sum over global products of $\sigma^{\pm}_l=\sigma^x_l \pm i \sigma^y_l$. Along with $\sigma^z_l$, these operators serve a basis to construct correlation functions of spins $\langle \Psi | \sigma^-_k(t)  \sigma^+_l(0) | \Psi \rangle$, for example. By judiciously encoding fermionic phase factors we then also evaluate expressions such as $\langle \Psi | c_k(t)  c^\dagger_l(0) | \Psi \rangle$. The synthesis techniques described above can be used to construct measurement gadgets for one- and two-point correlation functions---used in the natural sciences and high-rank polynomials moments as in the Mermin's entanglement witness. A comprehensive application of symmetric considerations to efficient entanglement classification and manipulation via measurement is left as future work. 

\section{Conclusions and Discussion} \label{sec:conclusion}
We have introduced and examined near term algorithms that utilize both product- and sum-formulas to sample symmetric operators on stochastic quantum, and deterministic classical computers, illustrating utility in constructing dynamical and stationary symmetric quantum operators. 

As a first illustrative application, we examine the stochastic dynamics of spectral random walks that preserve and break time-reversal-invariance. A second application demonstrates how our protocol can enhance the precision scaling of the first-order Trotter factorization of a time evolution operator from $\mathcal{O}(t^2)$ to $\mathcal{O}(t^3)$. We studied the case where a Hamiltonian may be decomposed into the $A$ and $B$ partitions and suggest the extension to fully-symmetrized $k$-term forms as an open direction, with, e.g., $\mathcal{O}(N^3)$ algorithms to prepare permutation $k$-term invariant  operators by quantum Schur transformation~\cite{Gu2021}.  
We expect our algorithm to both inspire future advanced unitary operator synthesis while also finding practical applications in the near term due to straightforward nature. In the third application, we present a symmetry-enhanced variational form within the context of preparing eigenstates of a spin-chain. Finally, our fourth application utilizes the intrinsic symmetries of Mermin-polynomials to provide a natural framework for efficient encoding and entanglement verification. 

In a sense, our algorithm can be viewed as a coherence-resource restricted, 
compared to coherent qubitization \cite{Low2019, Martyn2023}, model of quantum computation. Here, coherences, which are typically used to boost transition probabilities to unity (implicitly required to construct deterministic algorithms), are exchanged for accepting the outcomes of measurements. This approach represents a computationally restricted, yet more readily physically attainable, model of stochastic quantum computation. Despite this restriction, we nevertheless find a wide variety of applications including spectral quantum walks, time evolution, variational state preparation, and entanglement verification. We expect further investigations of probabilistic symmetric forms will reveal additional applications, bridging the gaps between theoretical realizations, available resources, and practical implementations. 

\begin{acknowledgments}
We thank Titus Morris for discussions and reviewing our manuscript. E.D.\ is supported by the U.S.\ Department of Energy, Office of Science, Advanced Scientific Research Program, Early Career Award under Contract No.~ERKJ420. S.C.\ acknowledges DOE ASCR funding under the Quantum Computing Application Teams program, FWP No.~ERKJ347. Y.W.\ is supported by the U.S.\ Department of Energy, Office of Science, National Quantum Information Science Research Centers, Quantum Science Center. This research used resources of the Oak Ridge Leadership Computing Facility, which is a DOE Office of Science User Facility supported under Contract No.~DE-AC05-00OR22725.
\end{acknowledgments}

\section*{Author Declarations}

\subsection*{Conflict of Interest}
The authors have no conflict of interest to disclose. 

\subsection*{Author Contributions}
\noindent
\textbf{Yan Wang:} Conceptualization (equal); Data curation (equal); Formal analysis (equal); Investigation (equal); Methodology (equal); Validation (equal); Visualization (equal); Writing -- original draft (equal); Writing -- review \& editing (equal).
\textbf{Sarah Chehade:} Conceptualization (equal); Data curation (equal); Formal analysis (equal); Investigation (equal); Methodology (equal); Validation (equal); Writing -- original draft (equal); Writing -- review \& editing (equal).
\textbf{Eugene Dumitrescu:} Conceptualization (equal); Data curation (equal); Formal analysis (equal); Investigation (equal); Methodology (equal); Validation (equal); Visualization (supporting); Writing -- original draft (equal); Writing -- review \& editing (equal).

\section*{Data Availability Statement}
The data that support the findings of this study are available from the corresponding authors upon reasonable request.

\appendix  

\section{Block-Encoding by the Gadget Circuit with One Ancilla Qubit} \label{sec:gadgetCircuitProof}
We derive the block-encoded nonunitary quantum operators shown in \cref{fig:circuit} and the scattered final state, given the initial state $\ket{0}_a \otimes \ket{\Psi}$ (similar for the initial state $\ket{1}_a \otimes \ket{\Psi}$). The following calculation is essentially the same as that for Lemma~2 (specifically $\kappa = 1$) in Ref.~\onlinecite{Childs2012}: 
\begin{align*}
 &\mathrel{\phantom{=}}
    (H_a \otimes \mathds{1})
      \Bigl(\op{0}_a \otimes \mathds{1} + \op{1}_a \otimes V \Bigr) 
      \Bigl(\op{0}_a \otimes U \\
 &\phantom{={}} + \op{1}_a \otimes \mathds{1} \Bigr)
      (H_a \otimes \mathds{1})  (\ket{0}_a \otimes \ket{\Psi}) \\
 &= \Bigl(\op{+}_a \otimes U + \op{-}_a \otimes V \Bigr)
      (\ket{0}_a \otimes \ket{\Psi}) \\
 &= \ket{+}\ip{+}{0}_a \otimes  U\ket{\Psi} + \ket{-}\ip{-}{0}_a \otimes  V\ket{\Psi} \\
 &= \Bigl( \op{0}_a \otimes \mathds{1} + \op{1}_a \otimes \mathds{1} \Bigr)
    \Bigl(\ket{+} \otimes \ip{+}{0}_a U\ket{\Psi} \\
 &\phantom{={}} + \ket{-} \otimes \ip{-}{0}_a V\ket{\Psi} \Bigr) \\
 &= \ket{0}_a \otimes 
      \Bigl(\ev{0 \op{+} 0}_a U + \ev{0 \op{-} 0}_a V \Bigr)\ket{\Psi} \\
 &\phantom{={}} +
      \ket{1}_a \otimes 
      \Bigl(\ev{1 \op{+} 0}_a U + \ev{1 \op{-} 0}_a V \Bigr)\ket{\Psi} 
 \\
 &= \ket{0}_a \otimes  \frac{U + V}{2}\ket{\Psi} + 
      \ket{1}_a \otimes  \frac{U - V}{2}\ket{\Psi} \\
 &\equiv
      \ket{0}_a \otimes  U_+\ket{\Psi} + 
      \ket{1}_a \otimes  U_-\ket{\Psi} .
\end{align*}
In the intermediate, step we used resolution of identity $\mathds{1}_a \otimes \mathds{1} = \op{0}_a \otimes \mathds{1} + \op{1}_a \otimes \mathds{1}$ in the final measurement basis of the ancilla.

\section{Gadget Circuit with One Continuous Variable Ancilla}
\label{sec:CV_decoherence}
Consider Fig.~\ref{fig:circuit} where the ancilla qubit is replaced by a continuous variable qumode. Denote the ancillary qumode controlled time evolution of the principal system as $e^{-i\hat{x}\otimes Ht}$, where $\hat{x}$ is a position operator for the qumode, and the initial state $\ket{0}\otimes \ket{\Psi}$, where $\ket{0}$ is the CV vacuum state and $\ket{\Psi} = \sum c_n \ket{n}$. The reduced density operator $\rho(t)$ upon tracing out the ancillary CV mode (i.e., measuring the ancillary but without ``recording data''---in practice, it means doing multiple shots and constructing an ensemble of the projected states) is
\begin{align}
     \rho(t) 
  &= \Tr_a \qty[e^{-i\hat{x}\otimes Ht} 
     \qty(\ket{0}\otimes \ket{\Psi} \bra{0}\otimes \bra{\Psi})
     e^{+i\hat{x}\otimes Ht}] \\
  &= \int_{-\infty}^{+\infty} dx \bra{x}
     \Big[e^{-i\hat{x}\otimes Ht} 
     (\ket{0}\otimes \ket{\Psi} \notag\\
  &\phantom{={}}\times \bra{0}\otimes \bra{\Psi})
     e^{+i\hat{x}\otimes Ht}\Big] \ket{x} \\
  &= \int_{-\infty}^{+\infty} dx\, \abs{\ip{x}{0}}^2
     e^{-ixHt} \op{\Psi}  e^{+ixHt} \\
  &= \sqrt{\frac{2}{\pi}} \int_{-\infty}^{+\infty} dx\, e^{-2x^2} 
    \Bigg( \sum_k |c_k|^2 \op{k} \notag\\
  &\phantom{={}} +
      \sum_{n\neq m} e^{-ix(\omega_n - \omega_m)t} c_n c_m^*\op{n}{m} \Bigg) \\
  &= \sum_k |c_k|^2 \op{k} + 
     \Bigg[\sqrt{\frac{2}{\pi}}\sum_{n\neq m}c_n c_m^*\op{n}{m} \notag\\
  &\phantom{={}}\times \int_{-\infty}^{+\infty} dx\,
     e^{-2x^2 -ix(\omega_n - \omega_m)t}\Bigg] \\
  &= \sum_k |c_k|^2 \op{k} \notag\\
  &\phantom{={}} + \sum_{n\neq m}e^{-t^2(\omega_n-\omega_m)^2/8} c_n c_m^*\op{n}{m} \\
  &\to \sum_k |c_k|^2 \op{k}
  \equiv \rho^\text{d}, \text{ as $t\to \infty$}.
\end{align}
By using a large time-step size $t$, we can obtain the decoherent diagonal ensemble similar to the result in \cref{sec:decoherence}.

\section{Compiled Circuit for Spectral-Projection Random Walk}
\label{sec:appRWcirc}
To compile the controlled time-evolution unitaries in \cref{fig:RWsimulations}(a) into a circuit suitable for NISQ hardware, we rewrite \cref{eq:RW_Ham_1q} into $H = U(\theta,\phi) H_d U^\dagger(\theta,\phi)$, where
\begin{align}
     H_d 
  &= \omega_{+} \mathds{1} + \omega_{-} Z, \\
     U(\theta,\phi) 
  &= e^{-i\frac{\phi}{2} Z} e^{-i\frac{\theta}{2} Y} 
  = R_z(\phi) R_y(\theta).
\end{align}
This can be obtained by plugging $\ket{g} = \cos\frac{\theta}{2}\ket{0} + e^{i\phi}\sin\frac{\theta}{2} \ket{1} = e^{i\phi/2}U(\theta, \phi) \ket{0}$ and $\ket{e} = \sin\frac{\theta}{2}\ket{0} - e^{i\phi}\cos\frac{\theta}{2} \ket{1} = - e^{i\phi/2} U(\theta, \phi) \ket{1} $ into $H = \omega_g \op{g} + \omega_e \op{e} = U(\theta,\phi) (\omega_g\op{0} + \omega_e\op{1}) U^\dagger(\theta,\phi)$ and using the fact $\op{0} = (\mathds{1}+Z)/2$ and $\op{1} = (\mathds{1}-Z)/2$.

Then, $e^{-itH} = U(\theta,\phi) e^{-itH_d} U^\dagger(\theta,\phi)$ and the two controlled unitary in \cref{fig:RWsimulations}(a) can be compiled in the following way:
\begin{align}
  &\mathrel{\phantom{=}}
     \op{0}_a \otimes e^{-itH} + 
     \op{1}_a \otimes e^{+itH} \notag \\
  &= e^{-it Z_a\otimes H} \\
  &= [\mathds{1}\otimes U(\theta,\phi)]
     e^{-it Z_a\otimes H_d}
     [\mathds{1}\otimes U^\dagger(\theta,\phi)] \\
  &= 
     \begin{aligned} \small
     \Qcircuit @C=0.4em @R=0.4em {%
       &\qw &\ctrl{1} &\gate{R_z(2\omega_+ t)} &\ctrl{1} &\qw &\qw \\
       &\gate{U^\dagger(\theta,\phi)} &\targ &\gate{R_z(2\omega_- t)} &\targ &\gate{U(\theta,\phi)} &\qw }
     \end{aligned}
\end{align}
where the $z$-rotation gate $R_z(\theta) = e^{-i\theta Z /2}$. For a random walk with $r$ steps of repetition of the above circuit, the $U$ gates in the middle cancel due to  $U(\theta,\phi) U^\dagger(\theta,\phi) = \mathds{1}$. In addition, for convenience, we can add a $U$ ($U^\dagger$) gate at the beginning (end) of the circuit so we can prepare initial state (measure the final state) in the eigenbasis $\{\ket{g}, \ket{e}\}$. Thus, no $U$ gates will be applied in our compiled circuit suitable for NISQ hardware. This is equivalent to simulate in the eigenbasis of $H$, but we stress our algorithm works in general setting if the controlled time-evolution oracle can be accurately implemented on the NISQ or fault-tolerant hardware. The final circuit used in numerical simulation is as follows:
\begin{align}
\begin{aligned} \small
  \Qcircuit @C=0.4em @R=0.4em {%
   & & & &\mbox{\hspace{2em} $r$ iterations} &\push{\raisebox{1.2em}{}}  & & & \\
   \lstick{\ket{0}_a}
   &\qw &\gate{H} &\ctrl{1} &\gate{R_{z}(2\omega_+ t)} &\ctrl{1} &\gate{H} &\meter  &\qw \\
   \lstick{\ket{0}}
   &\gate{R_y(\alpha)} &\qw &\targ &\gate{R_{z}(2\omega_- t)} &\targ &\qw &\ustick{m_\kappa}\qw &\mbox{$\; \ket{\Psi_{\kappa+1}}$} \gategroup{2}{3}{3}{8}{0.6em}{^\}}
  }
\end{aligned}    
\end{align}
Here, the computational basis of the system qubit (second row) is chosen to be the eigenbasis of $H$, and thus, $R_y(\alpha) = e^{-i\alpha Y/2}$ prepares an initial state $\cos\frac{\alpha}{2}\ket{g} + \sin\frac{\alpha}{2}\ket{e}$. We set $\alpha = \theta$ to prepare an initial state $\ket{\Psi_0}$ pointing to north pole of the Bloch sphere as shown in \cref{fig:RWsimulations}(a).

\section{Survival Probabilities for Quantum Zeno Effect}
\label{sec:appQZE}
As given in the main text, after $n$ measurements of ancilla with outcome $0$s [denote the success probability of obtaining these $0$ outcomes as $P(0)$], the final survival probability $S_n$ (conditioned on successfully measuring $0$s on ancilla), i.e., measurement of $\hat{O} = \op{\Psi}$ in the principal system at the final state $\ket{\Phi} = \cos^n(HT/n)\ket{\Psi}/ [\ev{\cos^{2n}(HT/n)}{\Psi}]^{1/2}$ (denote $HT = h$), is given by
\begin{align}
     S_n
  &\equiv S_n (\Phi\to \Psi \mid 0) \\
  &= \Tr(\op{\Phi}\hat{O})
   = \abs{\ip{\Psi}{\Phi}}^2 \\
  &= \frac{\abs{\ev{\cos^{n}(HT/n)}{\Psi}}^2}{\ev{\cos^{2n}(HT/n)}{\Psi}}\\
  &= \frac{\abs{\ev{1-\frac{h^2}{2n} + \frac{h^4}{8n^2} + \order{\frac{1}{n^3}}}{\Psi}}^2}
     {\ev{1-\frac{h^2}{n} + \frac{h^4}{2n^2} + \order{\frac{1}{n^3}}}{\Psi}} \\
  &= 1 - \frac{\overline{(h^2 - \overline{h^2})^2}}{4n^2} + \order{\frac{1}{n^3}},
\end{align}
where $\overline{\hat{o}} = \ev{\hat{o}}{\Psi}$. Since $P(0) = \ev{\cos^{2n}(HT/n)}{\Psi}$, the survival probability of the entire system is given by
\begin{align}
     \check{S}_n
  &\equiv S_n(\Phi\to \Psi \cap 0) \\
  &= S_n(\Phi\to \Psi \mid 0) P(0)
   = \abs{\ev{\cos^{n}(HT/n)}{\Psi}}^2
   \nonumber \\ &= 1 - \frac{\overline{h^2}}{n} + \order{\frac{1}{n^2}}. \nonumber
\end{align}
Note that this is different from unconditioned (i.e., measuring the ancilla but without post-selection) survival probability of the principal system alone $S_n (\Phi\to \Psi) = \sum_{k=0}^{2^n - 1} S_n(\Phi\to \Psi \mid k) P(k)$ (the binary representation of $k$ is the outcome string). Since $1 \geq S_n(\Phi\to \Psi) \geq  \check{S}_n \to 1$ at QZE measurement frequency limit, $P(0\mid \Phi\to \Psi) = \check{S}/S_n(\Phi\to \Psi)\to 1$, meaning that simply measuring the ancilla fast enough without recording the outcome, if the principal system is stationary at the end, then QZE indicates the ancilla is stationary (with high probability) during the entire process.

For a textbook case of quantum Zeno effect (QZE), the measurement $\hat{O}= \op{\Psi}$ is performed at each step, so the state is evolved over a period $T/n$ at each step, and the final survival probability is given by
\begin{align}
     \widetilde{S}_n 
  &= \widetilde{S}_1^n \\
  &= \abs{\mel{\Psi}{e^{-iHT/n}}{\Psi}}^{2n} \\
  &= 1 - \frac{\overline{(h-\overline{h})^2}}{n} + \order{\frac{1}{n^2}}.
\end{align}

Noting $0\leq \overline{(h-\overline{h})^2} = \overline{h^2} - \overline{h}^2 \leq \overline{h^2}$, we find $S_n \geq \widetilde{S}_n \geq \check{S}_n$. Higher survival rate means slower evolution under monitoring, i.e., stronger QZE. The principal part of our system shows stronger QZE than standard case, but when considered together with ancillary part, the result is similar to the standard case.

\bibliography{refs.bib}

\providecommand*\hyphen{-}
\begin{thebibliography}{57}%
\makeatletter
\providecommand \@ifxundefined [1]{%
 \@ifx{#1\undefined}
}%
\providecommand \@ifnum [1]{%
 \ifnum #1\expandafter \@firstoftwo
 \else \expandafter \@secondoftwo
 \fi
}%
\providecommand \@ifx [1]{%
 \ifx #1\expandafter \@firstoftwo
 \else \expandafter \@secondoftwo
 \fi
}%
\providecommand \natexlab [1]{#1}%
\providecommand \enquote  [1]{``#1''}%
\providecommand \bibnamefont  [1]{#1}%
\providecommand \bibfnamefont [1]{#1}%
\providecommand \citenamefont [1]{#1}%
\providecommand \href@noop [0]{\@secondoftwo}%
\providecommand \href [0]{\begingroup \@sanitize@url \@href}%
\providecommand \@href[1]{\@@startlink{#1}\@@href}%
\providecommand \@@href[1]{\endgroup#1\@@endlink}%
\providecommand \@sanitize@url [0]{\catcode `\\12\catcode `\$12\catcode
  `\&12\catcode `\#12\catcode `\^12\catcode `\_12\catcode `\%12\relax}%
\providecommand \@@startlink[1]{}%
\providecommand \@@endlink[0]{}%
\providecommand \url  [0]{\begingroup\@sanitize@url \@url }%
\providecommand \@url [1]{\endgroup\@href {#1}{\urlprefix }}%
\providecommand \urlprefix  [0]{URL }%
\providecommand \Eprint [0]{\href }%
\providecommand \doibase [0]{https://doi.org/}%
\providecommand \selectlanguage [0]{\@gobble}%
\providecommand \bibinfo  [0]{\@secondoftwo}%
\providecommand \bibfield  [0]{\@secondoftwo}%
\providecommand \translation [1]{[#1]}%
\providecommand \BibitemOpen [0]{}%
\providecommand \bibitemStop [0]{}%
\providecommand \bibitemNoStop [0]{.\EOS\space}%
\providecommand \EOS [0]{\spacefactor3000\relax}%
\providecommand \BibitemShut  [1]{\csname bibitem#1\endcsname}%
\let\auto@bib@innerbib\@empty
\bibitem [{\citenamefont {Childs}\ and\ \citenamefont
  {Wiebe}(2012)}]{Childs2012}%
  \BibitemOpen
  \bibfield  {author} {\bibinfo {author} {\bibfnamefont {A.~M.}\ \bibnamefont
  {Childs}}\ and\ \bibinfo {author} {\bibfnamefont {N.}~\bibnamefont {Wiebe}},\
  }\bibfield  {title} {\enquote {\bibinfo {title} {{Hamiltonian} simulation
  using linear combinations of unitary operations},}\ }\href
  {https://doi.org/10.26421/QIC12.11-12} {\bibfield  {journal} {\bibinfo
  {journal} {Quantum Info. Comput.}\ }\textbf {\bibinfo {volume} {12}},\
  \bibinfo {pages} {901--924} (\bibinfo {year} {2012})}\BibitemShut {NoStop}%
\bibitem [{\citenamefont {Low}\ and\ \citenamefont {Chuang}(2017)}]{Low2017}%
  \BibitemOpen
  \bibfield  {author} {\bibinfo {author} {\bibfnamefont {G.~H.}\ \bibnamefont
  {Low}}\ and\ \bibinfo {author} {\bibfnamefont {I.~L.}\ \bibnamefont
  {Chuang}},\ }\bibfield  {title} {\enquote {\bibinfo {title} {Optimal
  {H}amiltonian simulation by quantum signal processing},}\ }\href
  {https://doi.org/10.1103/PhysRevLett.118.010501} {\bibfield  {journal}
  {\bibinfo  {journal} {Phys. Rev. Lett.}\ }\textbf {\bibinfo {volume} {118}},\
  \bibinfo {pages} {010501} (\bibinfo {year} {2017})}\BibitemShut {NoStop}%
\bibitem [{\citenamefont {Rossi}\ \emph {et~al.}(2023)\citenamefont {Rossi},
  \citenamefont {Bastidas}, \citenamefont {Munro},\ and\ \citenamefont
  {Chuang}}]{Rossi2023}%
  \BibitemOpen
  \bibfield  {author} {\bibinfo {author} {\bibfnamefont {Z.~M.}\ \bibnamefont
  {Rossi}}, \bibinfo {author} {\bibfnamefont {V.~M.}\ \bibnamefont {Bastidas}},
  \bibinfo {author} {\bibfnamefont {W.~J.}\ \bibnamefont {Munro}},\ and\
  \bibinfo {author} {\bibfnamefont {I.~L.}\ \bibnamefont {Chuang}},\
  }\href@noop {} {\enquote {\bibinfo {title} {Quantum signal processing with
  continuous variables},}\ } (\bibinfo {year} {2023}),\ \Eprint
  {https://arxiv.org/abs/2304.14383} {arXiv:2304.14383 [quant-ph]} \BibitemShut
  {NoStop}%
\bibitem [{\citenamefont {Low}\ and\ \citenamefont {Chuang}(2019)}]{Low2019}%
  \BibitemOpen
  \bibfield  {author} {\bibinfo {author} {\bibfnamefont {G.~H.}\ \bibnamefont
  {Low}}\ and\ \bibinfo {author} {\bibfnamefont {I.~L.}\ \bibnamefont
  {Chuang}},\ }\bibfield  {title} {\enquote {\bibinfo {title} {{H}amiltonian
  simulation by qubitization},}\ }\href
  {https://doi.org/10.22331/q-2019-07-12-163} {\bibfield  {journal} {\bibinfo
  {journal} {Quantum}\ }\textbf {\bibinfo {volume} {3}},\ \bibinfo {pages}
  {163} (\bibinfo {year} {2019})}\BibitemShut {NoStop}%
\bibitem [{\citenamefont {Hu}, \citenamefont {Xia},\ and\ \citenamefont
  {Kais}(2020)}]{Hu2020}%
  \BibitemOpen
  \bibfield  {author} {\bibinfo {author} {\bibfnamefont {Z.}~\bibnamefont
  {Hu}}, \bibinfo {author} {\bibfnamefont {R.}~\bibnamefont {Xia}},\ and\
  \bibinfo {author} {\bibfnamefont {S.}~\bibnamefont {Kais}},\ }\bibfield
  {title} {\enquote {\bibinfo {title} {A quantum algorithm for evolving open
  quantum dynamics on quantum computing devices},}\ }\href
  {https://doi.org/10.1038/s41598-020-60321-x} {\bibfield  {journal} {\bibinfo
  {journal} {Sci. Rep.}\ }\textbf {\bibinfo {volume} {10}},\ \bibinfo {pages}
  {3301} (\bibinfo {year} {2020})}\BibitemShut {NoStop}%
\bibitem [{\citenamefont {Suri}\ \emph {et~al.}(2023)\citenamefont {Suri},
  \citenamefont {Barreto}, \citenamefont {Hadfield}, \citenamefont {Wiebe},
  \citenamefont {Wudarski},\ and\ \citenamefont {Marshall}}]{Suri2023}%
  \BibitemOpen
  \bibfield  {author} {\bibinfo {author} {\bibfnamefont {N.}~\bibnamefont
  {Suri}}, \bibinfo {author} {\bibfnamefont {J.}~\bibnamefont {Barreto}},
  \bibinfo {author} {\bibfnamefont {S.}~\bibnamefont {Hadfield}}, \bibinfo
  {author} {\bibfnamefont {N.}~\bibnamefont {Wiebe}}, \bibinfo {author}
  {\bibfnamefont {F.}~\bibnamefont {Wudarski}},\ and\ \bibinfo {author}
  {\bibfnamefont {J.}~\bibnamefont {Marshall}},\ }\bibfield  {title} {\enquote
  {\bibinfo {title} {Two-unitary decomposition algorithm and open quantum
  system simulation},}\ }\href {https://doi.org/10.22331/q-2023-05-15-1002}
  {\bibfield  {journal} {\bibinfo  {journal} {Quantum}\ }\textbf {\bibinfo
  {volume} {7}},\ \bibinfo {pages} {1002} (\bibinfo {year} {2023})}\BibitemShut
  {NoStop}%
\bibitem [{\citenamefont {Low}, \citenamefont {Kliuchnikov},\ and\
  \citenamefont {Wiebe}(2019)}]{Low2019a}%
  \BibitemOpen
  \bibfield  {author} {\bibinfo {author} {\bibfnamefont {G.~H.}\ \bibnamefont
  {Low}}, \bibinfo {author} {\bibfnamefont {V.}~\bibnamefont {Kliuchnikov}},\
  and\ \bibinfo {author} {\bibfnamefont {N.}~\bibnamefont {Wiebe}},\
  }\href@noop {} {\enquote {\bibinfo {title} {Well-conditioned multiproduct
  {H}amiltonian simulation},}\ } (\bibinfo {year} {2019}),\ \Eprint
  {https://arxiv.org/abs/1907.11679} {arXiv:1907.11679 [quant-ph]} \BibitemShut
  {NoStop}%
\bibitem [{\citenamefont {Carrera~Vazquez}\ \emph {et~al.}(2023)\citenamefont
  {Carrera~Vazquez}, \citenamefont {Egger}, \citenamefont {Ochsner},\ and\
  \citenamefont {Woerner}}]{CarreraVazquez2023}%
  \BibitemOpen
  \bibfield  {author} {\bibinfo {author} {\bibfnamefont {A.}~\bibnamefont
  {Carrera~Vazquez}}, \bibinfo {author} {\bibfnamefont {D.~J.}\ \bibnamefont
  {Egger}}, \bibinfo {author} {\bibfnamefont {D.}~\bibnamefont {Ochsner}},\
  and\ \bibinfo {author} {\bibfnamefont {S.}~\bibnamefont {Woerner}},\
  }\bibfield  {title} {\enquote {\bibinfo {title} {Well-conditioned
  multi-product formulas for hardware-friendly {H}amiltonian simulation},}\
  }\href {https://doi.org/10.22331/q-2023-07-25-1067} {\bibfield  {journal}
  {\bibinfo  {journal} {{Quantum}}\ }\textbf {\bibinfo {volume} {7}},\ \bibinfo
  {pages} {1067} (\bibinfo {year} {2023})}\BibitemShut {NoStop}%
\bibitem [{\citenamefont {Rendon}, \citenamefont {Watkins},\ and\ \citenamefont
  {Wiebe}(2022)}]{Rendon2022}%
  \BibitemOpen
  \bibfield  {author} {\bibinfo {author} {\bibfnamefont {G.}~\bibnamefont
  {Rendon}}, \bibinfo {author} {\bibfnamefont {J.}~\bibnamefont {Watkins}},\
  and\ \bibinfo {author} {\bibfnamefont {N.}~\bibnamefont {Wiebe}},\
  }\href@noop {} {\enquote {\bibinfo {title} {Improved accuracy for {T}rotter
  simulations using {C}hebyshev interpolation},}\ } (\bibinfo {year} {2022}),\
  \Eprint {https://arxiv.org/abs/2212.14144} {arXiv:2212.14144 [quant-ph]}
  \BibitemShut {NoStop}%
\bibitem [{\citenamefont {Zhuk}, \citenamefont {Robertson},\ and\ \citenamefont
  {Bravyi}(2023)}]{Zhuk2023}%
  \BibitemOpen
  \bibfield  {author} {\bibinfo {author} {\bibfnamefont {S.}~\bibnamefont
  {Zhuk}}, \bibinfo {author} {\bibfnamefont {N.}~\bibnamefont {Robertson}},\
  and\ \bibinfo {author} {\bibfnamefont {S.}~\bibnamefont {Bravyi}},\
  }\href@noop {} {\enquote {\bibinfo {title} {{T}rotter error bounds and
  dynamic multi-product formulas for {H}amiltonian simulation},}\ } (\bibinfo
  {year} {2023}),\ \Eprint {https://arxiv.org/abs/2306.12569} {arXiv:2306.12569
  [quant-ph]} \BibitemShut {NoStop}%
\bibitem [{\citenamefont {Martyn}\ \emph {et~al.}(2023)\citenamefont {Martyn},
  \citenamefont {Liu}, \citenamefont {Chin},\ and\ \citenamefont
  {Chuang}}]{Martyn2023}%
  \BibitemOpen
  \bibfield  {author} {\bibinfo {author} {\bibfnamefont {J.~M.}\ \bibnamefont
  {Martyn}}, \bibinfo {author} {\bibfnamefont {Y.}~\bibnamefont {Liu}},
  \bibinfo {author} {\bibfnamefont {Z.~E.}\ \bibnamefont {Chin}},\ and\
  \bibinfo {author} {\bibfnamefont {I.~L.}\ \bibnamefont {Chuang}},\ }\bibfield
   {title} {\enquote {\bibinfo {title} {Efficient fully-coherent quantum signal
  processing algorithms for real-time dynamics simulation},}\ }\href
  {https://doi.org/10.1063/5.0124385} {\bibfield  {journal} {\bibinfo
  {journal} {J. Chem. Phys.}\ }\textbf {\bibinfo {volume} {158}},\ \bibinfo
  {pages} {024106} (\bibinfo {year} {2023})}\BibitemShut {NoStop}%
\bibitem [{\citenamefont {Trotter}(1959)}]{Trotter1959}%
  \BibitemOpen
  \bibfield  {author} {\bibinfo {author} {\bibfnamefont {H.~F.}\ \bibnamefont
  {Trotter}},\ }\bibfield  {title} {\enquote {\bibinfo {title} {On the product
  of semi-groups of operators},}\ }\href {https://doi.org/10.2307/2033649}
  {\bibfield  {journal} {\bibinfo  {journal} {Proc. Amer. Math. Soc.}\ }\textbf
  {\bibinfo {volume} {10}},\ \bibinfo {pages} {545--551} (\bibinfo {year}
  {1959})}\BibitemShut {NoStop}%
\bibitem [{\citenamefont {Suzuki}(1976)}]{Suzuki1976}%
  \BibitemOpen
  \bibfield  {author} {\bibinfo {author} {\bibfnamefont {M.}~\bibnamefont
  {Suzuki}},\ }\bibfield  {title} {\enquote {\bibinfo {title} {Generalized
  {T}rotter's formula and systematic approximants of exponential operators and
  inner derivations with applications to many-body problems},}\ }\href
  {https://doi.org/10.1007/BF01609348} {\bibfield  {journal} {\bibinfo
  {journal} {Commun. Math. Phys.}\ }\textbf {\bibinfo {volume} {51}},\ \bibinfo
  {pages} {183--190} (\bibinfo {year} {1976})}\BibitemShut {NoStop}%
\bibitem [{\citenamefont {Childs}\ \emph {et~al.}(2021)\citenamefont {Childs},
  \citenamefont {Su}, \citenamefont {Tran}, \citenamefont {Wiebe},\ and\
  \citenamefont {Zhu}}]{Childs2021}%
  \BibitemOpen
  \bibfield  {author} {\bibinfo {author} {\bibfnamefont {A.~M.}\ \bibnamefont
  {Childs}}, \bibinfo {author} {\bibfnamefont {Y.}~\bibnamefont {Su}}, \bibinfo
  {author} {\bibfnamefont {M.~C.}\ \bibnamefont {Tran}}, \bibinfo {author}
  {\bibfnamefont {N.}~\bibnamefont {Wiebe}},\ and\ \bibinfo {author}
  {\bibfnamefont {S.}~\bibnamefont {Zhu}},\ }\bibfield  {title} {\enquote
  {\bibinfo {title} {Theory of {T}rotter error with commutator scaling},}\
  }\href {https://doi.org/10.1103/PhysRevX.11.011020} {\bibfield  {journal}
  {\bibinfo  {journal} {Phys. Rev. X}\ }\textbf {\bibinfo {volume} {11}},\
  \bibinfo {pages} {011020} (\bibinfo {year} {2021})}\BibitemShut {NoStop}%
\bibitem [{\citenamefont {Haah}\ \emph {et~al.}(2021)\citenamefont {Haah},
  \citenamefont {Hastings}, \citenamefont {Kothari},\ and\ \citenamefont
  {Low}}]{Haah2021}%
  \BibitemOpen
  \bibfield  {author} {\bibinfo {author} {\bibfnamefont {J.}~\bibnamefont
  {Haah}}, \bibinfo {author} {\bibfnamefont {M.~B.}\ \bibnamefont {Hastings}},
  \bibinfo {author} {\bibfnamefont {R.}~\bibnamefont {Kothari}},\ and\ \bibinfo
  {author} {\bibfnamefont {G.~H.}\ \bibnamefont {Low}},\ }\bibfield  {title}
  {\enquote {\bibinfo {title} {Quantum algorithm for simulating real time
  evolution of lattice {H}amiltonians},}\ }\href
  {https://doi.org/10.1137/18M1231511} {\bibfield  {journal} {\bibinfo
  {journal} {SIAM J. Comput.}\ }\textbf {\bibinfo {volume} {0}},\ \bibinfo
  {pages} {FOCS18\hyphen250--FOCS18\hyphen284} (\bibinfo {year}
  {2021})}\BibitemShut {NoStop}%
\bibitem [{\citenamefont {Ostmeyer}(2023)}]{Ostmeyer2023}%
  \BibitemOpen
  \bibfield  {author} {\bibinfo {author} {\bibfnamefont {J.}~\bibnamefont
  {Ostmeyer}},\ }\bibfield  {title} {\enquote {\bibinfo {title} {Optimised
  {T}rotter decompositions for classical and quantum computing},}\ }\href
  {https://doi.org/10.1088/1751-8121/acde7a} {\bibfield  {journal} {\bibinfo
  {journal} {J. Phys. A: Math. Theor.}\ }\textbf {\bibinfo {volume} {56}},\
  \bibinfo {pages} {285303} (\bibinfo {year} {2023})}\BibitemShut {NoStop}%
\bibitem [{\citenamefont {Zeng}\ \emph {et~al.}(2022)\citenamefont {Zeng},
  \citenamefont {Sun}, \citenamefont {Jiang},\ and\ \citenamefont
  {Zhao}}]{Zeng2022}%
  \BibitemOpen
  \bibfield  {author} {\bibinfo {author} {\bibfnamefont {P.}~\bibnamefont
  {Zeng}}, \bibinfo {author} {\bibfnamefont {J.}~\bibnamefont {Sun}}, \bibinfo
  {author} {\bibfnamefont {L.}~\bibnamefont {Jiang}},\ and\ \bibinfo {author}
  {\bibfnamefont {Q.}~\bibnamefont {Zhao}},\ }\href@noop {} {\enquote {\bibinfo
  {title} {Simple and high-precision {H}amiltonian simulation by compensating
  {T}rotter error with linear combination of unitary operations},}\ } (\bibinfo
  {year} {2022}),\ \Eprint {https://arxiv.org/abs/2212.04566} {arXiv:2212.04566
  [quant-ph]} \BibitemShut {NoStop}%
\bibitem [{Note1()}]{Note1}%
  \BibitemOpen
  \bibinfo {note} {We use the term ``integrator'' for the propagator for both
  classical Hamiltonians~\cite {Yoshida1990} and quantum Hamiltonians. In the
  latter case, it becomes the time-evolution operator.}\BibitemShut {Stop}%
\bibitem [{\citenamefont {Suzuki}(1991)}]{Suzuki1991}%
  \BibitemOpen
  \bibfield  {author} {\bibinfo {author} {\bibfnamefont {M.}~\bibnamefont
  {Suzuki}},\ }\bibfield  {title} {\enquote {\bibinfo {title} {General theory
  of fractal path integrals with applications to many‐body theories and
  statistical physics},}\ }\href {https://doi.org/10.1063/1.529425} {\bibfield
  {journal} {\bibinfo  {journal} {J. Math. Phys.}\ }\textbf {\bibinfo {volume}
  {32}},\ \bibinfo {pages} {400--407} (\bibinfo {year} {1991})}\BibitemShut
  {NoStop}%
\bibitem [{Note2()}]{Note2}%
  \BibitemOpen
  \bibinfo {note} {This includes both symmetry and anti-symmetry, associated
  with $+1$ and $-1$ eigenvalue, respectively.}\BibitemShut {Stop}%
\bibitem [{\citenamefont {Jordan}(1932)}]{Jordan1932}%
  \BibitemOpen
  \bibfield  {author} {\bibinfo {author} {\bibfnamefont {P.}~\bibnamefont
  {Jordan}},\ }\bibfield  {title} {\enquote {\bibinfo {title} {{{\"U}ber eine
  Klasse nichtassoziativer hyperkomplexer Algebren}},}\ }\href
  {http://eudml.org/doc/59403} {\bibfield  {journal} {\bibinfo  {journal}
  {{Nachrichten von der Gesellschaft der Wissenschaften zu G{\"o}ttingen,
  Mathematisch-Physikalische Klasse}}\ }\textbf {\bibinfo {volume} {1932}},\
  \bibinfo {pages} {569--575} (\bibinfo {year} {1932})}\BibitemShut {NoStop}%
\bibitem [{\citenamefont {Magnus}(1954)}]{magnus1954exponential}%
  \BibitemOpen
  \bibfield  {author} {\bibinfo {author} {\bibfnamefont {W.}~\bibnamefont
  {Magnus}},\ }\bibfield  {title} {\enquote {\bibinfo {title} {On the
  exponential solution of differential equations for a linear operator},}\
  }\href@noop {} {\bibfield  {journal} {\bibinfo  {journal} {Communications on
  pure and applied mathematics}\ }\textbf {\bibinfo {volume} {7}},\ \bibinfo
  {pages} {649--673} (\bibinfo {year} {1954})}\BibitemShut {NoStop}%
\bibitem [{\citenamefont {Goldberg}(1956)}]{Goldberg1956}%
  \BibitemOpen
  \bibfield  {author} {\bibinfo {author} {\bibfnamefont {K.}~\bibnamefont
  {Goldberg}},\ }\bibfield  {title} {\enquote {\bibinfo {title} {The formal
  power series for $\log e^{x} e^{y}$},}\ }\href
  {https://doi.org/10.1215/S0012-7094-56-02302-X} {\bibfield  {journal}
  {\bibinfo  {journal} {Duke Math. J.}\ }\textbf {\bibinfo {volume} {23}},\
  \bibinfo {pages} {13--21} (\bibinfo {year} {1956})}\BibitemShut {NoStop}%
\bibitem [{\citenamefont {Bonfiglioli}\ and\ \citenamefont
  {Fulci}(2012)}]{Bonfiglioli2012}%
  \BibitemOpen
  \bibfield  {author} {\bibinfo {author} {\bibfnamefont {A.}~\bibnamefont
  {Bonfiglioli}}\ and\ \bibinfo {author} {\bibfnamefont {R.}~\bibnamefont
  {Fulci}},\ }\href {https://doi.org/10.1007/978-3-642-22597-0} {\emph
  {\bibinfo {title} {Topics in Noncommutative Algebra: The Theorem of
  {C}ampbell, {B}aker, {H}ausdorff and {D}ynkin}}},\ Lecture Notes in
  Mathematics\ (\bibinfo  {publisher} {Springer Berlin, Heidelberg},\ \bibinfo
  {year} {2012})\BibitemShut {NoStop}%
\bibitem [{\citenamefont {Chehade}, \citenamefont {Wang},\ and\ \citenamefont
  {Wang}(2024)}]{chehade2024suzuki}%
  \BibitemOpen
  \bibfield  {author} {\bibinfo {author} {\bibfnamefont {S.}~\bibnamefont
  {Chehade}}, \bibinfo {author} {\bibfnamefont {S.}~\bibnamefont {Wang}},\ and\
  \bibinfo {author} {\bibfnamefont {Z.}~\bibnamefont {Wang}},\ }\bibfield
  {title} {\enquote {\bibinfo {title} {{S}uzuki type estimates for
  exponentiated sums and generalized {L}ie-{T}rotter formulas in
  {JB}-algebras},}\ }\href {https://doi.org/10.1016/j.laa.2023.10.004}
  {\bibfield  {journal} {\bibinfo  {journal} {Linear Algebra Appl.}\ }\textbf
  {\bibinfo {volume} {680}},\ \bibinfo {pages} {156--169} (\bibinfo {year}
  {2024})}\BibitemShut {NoStop}%
\bibitem [{\citenamefont {Dynkin}(2000)}]{Dynkin2000}%
  \BibitemOpen
  \bibfield  {author} {\bibinfo {author} {\bibfnamefont {E.~B.}\ \bibnamefont
  {Dynkin}},\ }\enquote {\bibinfo {title} {Selected papers of {E.\ B.\ Dynkin}
  with commentary},}\ \ (\bibinfo  {publisher} {American Mathematical
  Society},\ \bibinfo {year} {2000})\ Chap.\ \bibinfo {chapter} {Calculation of
  the coefficients in the {C}ampbell-{H}ausdorff formula}, pp.\ \bibinfo
  {pages} {31--35},\ \bibinfo {note} {originally published (in Russian) in
  Doklady Akad. Nauk SSSR 57, 323--326 (1947).}\BibitemShut {Stop}%
\bibitem [{\citenamefont {Monras}, \citenamefont {Beige},\ and\ \citenamefont
  {Wiesner}(2011)}]{Monras2011}%
  \BibitemOpen
  \bibfield  {author} {\bibinfo {author} {\bibfnamefont {A.}~\bibnamefont
  {Monras}}, \bibinfo {author} {\bibfnamefont {A.}~\bibnamefont {Beige}},\ and\
  \bibinfo {author} {\bibfnamefont {K.}~\bibnamefont {Wiesner}},\ }\href@noop
  {} {\enquote {\bibinfo {title} {Hidden quantum {M}arkov models and
  non-adaptive read-out of many-body states},}\ } (\bibinfo {year} {2011}),\
  \Eprint {https://arxiv.org/abs/1002.2337} {arXiv:1002.2337 [quant-ph]}
  \BibitemShut {NoStop}%
\bibitem [{\citenamefont {Milz}\ and\ \citenamefont {Modi}(2021)}]{Milz2021}%
  \BibitemOpen
  \bibfield  {author} {\bibinfo {author} {\bibfnamefont {S.}~\bibnamefont
  {Milz}}\ and\ \bibinfo {author} {\bibfnamefont {K.}~\bibnamefont {Modi}},\
  }\bibfield  {title} {\enquote {\bibinfo {title} {Quantum stochastic processes
  and quantum non-{M}arkovian phenomena},}\ }\href
  {https://doi.org/10.1103/PRXQuantum.2.030201} {\bibfield  {journal} {\bibinfo
   {journal} {PRX Quantum}\ }\textbf {\bibinfo {volume} {2}},\ \bibinfo {pages}
  {030201} (\bibinfo {year} {2021})}\BibitemShut {NoStop}%
\bibitem [{\citenamefont {Lacroix}, \citenamefont {Ruiz~Guzman},\ and\
  \citenamefont {Siwach}(2023)}]{lacroix2023}%
  \BibitemOpen
  \bibfield  {author} {\bibinfo {author} {\bibfnamefont {D.}~\bibnamefont
  {Lacroix}}, \bibinfo {author} {\bibfnamefont {E.~A.}\ \bibnamefont
  {Ruiz~Guzman}},\ and\ \bibinfo {author} {\bibfnamefont {P.}~\bibnamefont
  {Siwach}},\ }\bibfield  {title} {\enquote {\bibinfo {title} {Symmetry
  breaking/symmetry preserving circuits and symmetry restoration on quantum
  computers},}\ }\href {https://doi.org/10.1140/epja/s10050-022-00911-7}
  {\bibfield  {journal} {\bibinfo  {journal} {The European Physical Journal A}\
  }\textbf {\bibinfo {volume} {59}},\ \bibinfo {pages} {3} (\bibinfo {year}
  {2023})}\BibitemShut {NoStop}%
\bibitem [{\citenamefont {Chen}\ and\ \citenamefont {Wei}(2020)}]{Chen2020}%
  \BibitemOpen
  \bibfield  {author} {\bibinfo {author} {\bibfnamefont {Y.}~\bibnamefont
  {Chen}}\ and\ \bibinfo {author} {\bibfnamefont {T.-C.}\ \bibnamefont {Wei}},\
  }\bibfield  {title} {\enquote {\bibinfo {title} {Quantum algorithm for
  spectral projection by measuring an ancilla iteratively},}\ }\href
  {https://doi.org/10.1103/PhysRevA.101.032339} {\bibfield  {journal} {\bibinfo
   {journal} {Phys. Rev. A}\ }\textbf {\bibinfo {volume} {101}},\ \bibinfo
  {pages} {032339} (\bibinfo {year} {2020})}\BibitemShut {NoStop}%
\bibitem [{\citenamefont {Granade}\ and\ \citenamefont
  {Wiebe}(2022)}]{Granade2022}%
  \BibitemOpen
  \bibfield  {author} {\bibinfo {author} {\bibfnamefont {C.}~\bibnamefont
  {Granade}}\ and\ \bibinfo {author} {\bibfnamefont {N.}~\bibnamefont
  {Wiebe}},\ }\href@noop {} {\enquote {\bibinfo {title} {Using random walks for
  iterative phase estimation},}\ } (\bibinfo {year} {2022}),\ \Eprint
  {https://arxiv.org/abs/2208.04526} {arXiv:2208.04526 [quant-ph]} \BibitemShut
  {NoStop}%
\bibitem [{\citenamefont {Ferris}\ \emph {et~al.}(2023)\citenamefont {Ferris},
  \citenamefont {Wang}, \citenamefont {Hen}, \citenamefont {Kalev},
  \citenamefont {Bronn},\ and\ \citenamefont {Vlcek}}]{Ferris2023}%
  \BibitemOpen
  \bibfield  {author} {\bibinfo {author} {\bibfnamefont {K.~J.}\ \bibnamefont
  {Ferris}}, \bibinfo {author} {\bibfnamefont {Z.}~\bibnamefont {Wang}},
  \bibinfo {author} {\bibfnamefont {I.}~\bibnamefont {Hen}}, \bibinfo {author}
  {\bibfnamefont {A.}~\bibnamefont {Kalev}}, \bibinfo {author} {\bibfnamefont
  {N.~T.}\ \bibnamefont {Bronn}},\ and\ \bibinfo {author} {\bibfnamefont
  {V.}~\bibnamefont {Vlcek}},\ }\href@noop {} {\enquote {\bibinfo {title}
  {Exploiting maximally mixed states for spectral estimation by time
  evolution},}\ } (\bibinfo {year} {2023}),\ \Eprint
  {https://arxiv.org/abs/2312.00687} {arXiv:2312.00687 [quant-ph]} \BibitemShut
  {NoStop}%
\bibitem [{\citenamefont {Choi}\ \emph {et~al.}(2021)\citenamefont {Choi},
  \citenamefont {Lee}, \citenamefont {Bonitati}, \citenamefont {Qian},\ and\
  \citenamefont {Watkins}}]{Choi2021}%
  \BibitemOpen
  \bibfield  {author} {\bibinfo {author} {\bibfnamefont {K.}~\bibnamefont
  {Choi}}, \bibinfo {author} {\bibfnamefont {D.}~\bibnamefont {Lee}}, \bibinfo
  {author} {\bibfnamefont {J.}~\bibnamefont {Bonitati}}, \bibinfo {author}
  {\bibfnamefont {Z.}~\bibnamefont {Qian}},\ and\ \bibinfo {author}
  {\bibfnamefont {J.}~\bibnamefont {Watkins}},\ }\bibfield  {title} {\enquote
  {\bibinfo {title} {Rodeo algorithm for quantum computing},}\ }\href
  {https://doi.org/10.1103/PhysRevLett.127.040505} {\bibfield  {journal}
  {\bibinfo  {journal} {Phys. Rev. Lett.}\ }\textbf {\bibinfo {volume} {127}},\
  \bibinfo {pages} {040505} (\bibinfo {year} {2021})}\BibitemShut {NoStop}%
\bibitem [{\citenamefont {Stetcu}, \citenamefont {Baroni},\ and\ \citenamefont
  {Carlson}(2023)}]{Stetcu2023}%
  \BibitemOpen
  \bibfield  {author} {\bibinfo {author} {\bibfnamefont {I.}~\bibnamefont
  {Stetcu}}, \bibinfo {author} {\bibfnamefont {A.}~\bibnamefont {Baroni}},\
  and\ \bibinfo {author} {\bibfnamefont {J.}~\bibnamefont {Carlson}},\
  }\bibfield  {title} {\enquote {\bibinfo {title} {Projection algorithm for
  state preparation on quantum computers},}\ }\href
  {https://doi.org/10.1103/PhysRevC.108.L031306} {\bibfield  {journal}
  {\bibinfo  {journal} {Phys. Rev. C}\ }\textbf {\bibinfo {volume} {108}},\
  \bibinfo {pages} {L031306} (\bibinfo {year} {2023})}\BibitemShut {NoStop}%
\bibitem [{\citenamefont {Novotn{\'{y}}}, \citenamefont {Alber},\ and\
  \citenamefont {Jex}(2009)}]{Novotny2009}%
  \BibitemOpen
  \bibfield  {author} {\bibinfo {author} {\bibfnamefont {J.}~\bibnamefont
  {Novotn{\'{y}}}}, \bibinfo {author} {\bibfnamefont {G.}~\bibnamefont
  {Alber}},\ and\ \bibinfo {author} {\bibfnamefont {I.}~\bibnamefont {Jex}},\
  }\bibfield  {title} {\enquote {\bibinfo {title} {Random unitary dynamics of
  quantum networks},}\ }\href {https://doi.org/10.1088/1751-8113/42/28/282003}
  {\bibfield  {journal} {\bibinfo  {journal} {J. Phys. A: Math. Theor.}\
  }\textbf {\bibinfo {volume} {42}},\ \bibinfo {pages} {282003} (\bibinfo
  {year} {2009})}\BibitemShut {NoStop}%
\bibitem [{Note3()}]{Note3}%
  \BibitemOpen
  \bibinfo {note} {Should there be any degeneracy, a unitary transformation
  within the degenerate subspace can put those terms into the $\rho ^\protect
  \text {d}$ part, while all new basis states are still energy eigenstates. In
  the case of pure-state density operator $\op {\Psi }$ with $\ket {\Psi } =
  \DOTSB \sum@ \slimits@ _n c_n \ket {n}$, this transformation can be simply
  done by grouping all degenerate eigenstates corresponding to the same
  eigenenergy into a new eigenstate with a proper normalization.}\BibitemShut
  {Stop}%
\bibitem [{\citenamefont {Levin}\ and\ \citenamefont
  {Peres}(2017)}]{Levin2017}%
  \BibitemOpen
  \bibfield  {author} {\bibinfo {author} {\bibfnamefont {D.~A.}\ \bibnamefont
  {Levin}}\ and\ \bibinfo {author} {\bibfnamefont {Y.}~\bibnamefont {Peres}},\
  }\href@noop {} {\emph {\bibinfo {title} {{M}arkov Chains and Mixing Times}}}\
  (\bibinfo  {publisher} {American Mathematical Society, Providence, Rhode
  Island},\ \bibinfo {year} {2017})\BibitemShut {NoStop}%
\bibitem [{\citenamefont {K\'alm\'an}\ and\ \citenamefont
  {Kiss}(2018)}]{Kalman2018}%
  \BibitemOpen
  \bibfield  {author} {\bibinfo {author} {\bibfnamefont {O.}~\bibnamefont
  {K\'alm\'an}}\ and\ \bibinfo {author} {\bibfnamefont {T.}~\bibnamefont
  {Kiss}},\ }\bibfield  {title} {\enquote {\bibinfo {title} {Quantum state
  matching of qubits via measurement-induced nonlinear transformations},}\
  }\href {https://doi.org/10.1103/PhysRevA.97.032125} {\bibfield  {journal}
  {\bibinfo  {journal} {Phys. Rev. A}\ }\textbf {\bibinfo {volume} {97}},\
  \bibinfo {pages} {032125} (\bibinfo {year} {2018})}\BibitemShut {NoStop}%
\bibitem [{\citenamefont {Ambroladze}\ and\ \citenamefont
  {Wallin}(2000)}]{Ambroladze2000}%
  \BibitemOpen
  \bibfield  {author} {\bibinfo {author} {\bibfnamefont {A.~M. I. R. A.~N.}\
  \bibnamefont {Ambroladze}}\ and\ \bibinfo {author} {\bibfnamefont {H.~A.
  N.~S.}\ \bibnamefont {Wallin}},\ }\bibfield  {title} {\enquote {\bibinfo
  {title} {Random iteration of {M}\"{o}bius transformations and {F}urstenberg's
  theorem},}\ }\href {https://doi.org/10.1017/S0143385700000535} {\bibfield
  {journal} {\bibinfo  {journal} {Ergodic Theory Dynam. Systems}\ }\textbf
  {\bibinfo {volume} {20}},\ \bibinfo {pages} {953--962} (\bibinfo {year}
  {2000})}\BibitemShut {NoStop}%
\bibitem [{\citenamefont {McCarthy}\ \emph {et~al.}(2018)\citenamefont
  {McCarthy}, \citenamefont {Nop}, \citenamefont {Rastegar},\ and\
  \citenamefont {Roitershtein}}]{McCarthy2018}%
  \BibitemOpen
  \bibfield  {author} {\bibinfo {author} {\bibfnamefont {C.}~\bibnamefont
  {McCarthy}}, \bibinfo {author} {\bibfnamefont {G.}~\bibnamefont {Nop}},
  \bibinfo {author} {\bibfnamefont {R.}~\bibnamefont {Rastegar}},\ and\
  \bibinfo {author} {\bibfnamefont {A.}~\bibnamefont {Roitershtein}},\
  }\href@noop {} {\enquote {\bibinfo {title} {Random walk on the {P}oincar\'{e}
  disk induced by a group of {M}\"{o}bius transformations},}\ } (\bibinfo
  {year} {2018}),\ \Eprint {https://arxiv.org/abs/1804.06263} {arXiv:1804.06263
  [math.PR]} \BibitemShut {NoStop}%
\bibitem [{\citenamefont {Ge}, \citenamefont {Tura},\ and\ \citenamefont
  {Cirac}(2019)}]{Ge2019}%
  \BibitemOpen
  \bibfield  {author} {\bibinfo {author} {\bibfnamefont {Y.}~\bibnamefont
  {Ge}}, \bibinfo {author} {\bibfnamefont {J.}~\bibnamefont {Tura}},\ and\
  \bibinfo {author} {\bibfnamefont {J.~I.}\ \bibnamefont {Cirac}},\ }\bibfield
  {title} {\enquote {\bibinfo {title} {Faster ground state preparation and
  high-precision ground energy estimation with fewer qubits},}\ }\href
  {https://doi.org/10.1063/1.5027484} {\bibfield  {journal} {\bibinfo
  {journal} {Journal of Mathematical Physics}\ }\textbf {\bibinfo {volume}
  {60}} (\bibinfo {year} {2019}),\ 10.1063/1.5027484}\BibitemShut {NoStop}%
\bibitem [{\citenamefont {Lin}\ and\ \citenamefont {Tong}(2020)}]{Lin2020}%
  \BibitemOpen
  \bibfield  {author} {\bibinfo {author} {\bibfnamefont {L.}~\bibnamefont
  {Lin}}\ and\ \bibinfo {author} {\bibfnamefont {Y.}~\bibnamefont {Tong}},\
  }\bibfield  {title} {\enquote {\bibinfo {title} {Near-optimal ground state
  preparation},}\ }\href {https://doi.org/10.22331/Q-2020-12-14-372} {\bibfield
   {journal} {\bibinfo  {journal} {Quantum}\ }\textbf {\bibinfo {volume} {4}},\
  \bibinfo {pages} {372} (\bibinfo {year} {2020})}\BibitemShut {NoStop}%
\bibitem [{\citenamefont {Keen}, \citenamefont {Dumitrescu},\ and\
  \citenamefont {Wang}(2021)}]{Keen2021}%
  \BibitemOpen
  \bibfield  {author} {\bibinfo {author} {\bibfnamefont {T.}~\bibnamefont
  {Keen}}, \bibinfo {author} {\bibfnamefont {E.}~\bibnamefont {Dumitrescu}},\
  and\ \bibinfo {author} {\bibfnamefont {Y.}~\bibnamefont {Wang}},\ }\href
  {http://arxiv.org/abs/2112.05731} {\enquote {\bibinfo {title} {Quantum
  algorithms for ground-state preparation and {G}reen's function
  calculation},}\ } (\bibinfo {year} {2021}),\ \Eprint
  {https://arxiv.org/abs/2112.05731} {arXiv:2112.05731} \BibitemShut {NoStop}%
\bibitem [{\citenamefont {Hatano}\ and\ \citenamefont
  {Suzuki}(2005)}]{Hatano2005}%
  \BibitemOpen
  \bibfield  {author} {\bibinfo {author} {\bibfnamefont {N.}~\bibnamefont
  {Hatano}}\ and\ \bibinfo {author} {\bibfnamefont {M.}~\bibnamefont
  {Suzuki}},\ }\bibfield  {title} {\enquote {\bibinfo {title} {Finding
  exponential product formulas of higher orders},}\ }in\ \href
  {https://doi.org/10.1007/11526216_2} {\emph {\bibinfo {booktitle} {Quantum
  Annealing and Other Optimization Methods}}},\ \bibinfo {editor} {edited by\
  \bibinfo {editor} {\bibfnamefont {A.}~\bibnamefont {Das}}\ and\ \bibinfo
  {editor} {\bibfnamefont {B.}~\bibnamefont {K.~Chakrabarti}}}\ (\bibinfo
  {publisher} {Springer Berlin Heidelberg},\ \bibinfo {address} {Berlin,
  Heidelberg},\ \bibinfo {year} {2005})\ pp.\ \bibinfo {pages}
  {37--68}\BibitemShut {NoStop}%
\bibitem [{Note4()}]{Note4}%
  \BibitemOpen
  \bibinfo {note} {The Frobenius norm is defined as $\norm {A}_\protect \text
  {F} = [\Tr (A^\dagger A)]^{1/2}$. Note that the operator norm is an upper
  bound of the Trotter error for state time evolution $\norm {\delta U \ket
  {\Psi }} \leq \norm {\delta U} \norm {\ket {\Psi }} = \norm {\delta U}$,
  assuming $\ket {\Psi }$ is normalized.}\BibitemShut {Stop}%
\bibitem [{\citenamefont {Watrous}(2008)}]{Watrous2008}%
  \BibitemOpen
  \bibfield  {author} {\bibinfo {author} {\bibfnamefont {J.}~\bibnamefont
  {Watrous}},\ }\href@noop {} {\enquote {\bibinfo {title} {Quantum
  computational complexity},}\ } (\bibinfo {year} {2008}),\ \Eprint
  {https://arxiv.org/abs/0804.3401} {arXiv:0804.3401 [quant-ph]} \BibitemShut
  {NoStop}%
\bibitem [{\citenamefont {Schuch}\ and\ \citenamefont
  {Verstraete}(2009)}]{Schuch2009}%
  \BibitemOpen
  \bibfield  {author} {\bibinfo {author} {\bibfnamefont {N.}~\bibnamefont
  {Schuch}}\ and\ \bibinfo {author} {\bibfnamefont {F.}~\bibnamefont
  {Verstraete}},\ }\bibfield  {title} {\enquote {\bibinfo {title}
  {Computational complexity of interacting electrons and fundamental
  limitations of density functional theory},}\ }\href
  {https://doi.org/10.1038/nphys1370} {\bibfield  {journal} {\bibinfo
  {journal} {Nat. Phys.}\ }\textbf {\bibinfo {volume} {5}},\ \bibinfo {pages}
  {732--735} (\bibinfo {year} {2009})}\BibitemShut {NoStop}%
\bibitem [{\citenamefont {Wecker}, \citenamefont {Hastings},\ and\
  \citenamefont {Troyer}(2015)}]{Wecker2015}%
  \BibitemOpen
  \bibfield  {author} {\bibinfo {author} {\bibfnamefont {D.}~\bibnamefont
  {Wecker}}, \bibinfo {author} {\bibfnamefont {M.~B.}\ \bibnamefont
  {Hastings}},\ and\ \bibinfo {author} {\bibfnamefont {M.}~\bibnamefont
  {Troyer}},\ }\bibfield  {title} {\enquote {\bibinfo {title} {Progress towards
  practical quantum variational algorithms},}\ }\href
  {https://doi.org/10.1103/PhysRevA.92.042303} {\bibfield  {journal} {\bibinfo
  {journal} {Phys. Rev. A}\ }\textbf {\bibinfo {volume} {92}},\ \bibinfo
  {pages} {042303} (\bibinfo {year} {2015})}\BibitemShut {NoStop}%
\bibitem [{\citenamefont {Seki}, \citenamefont {Shirakawa},\ and\ \citenamefont
  {Yunoki}(2020)}]{Seki2020}%
  \BibitemOpen
  \bibfield  {author} {\bibinfo {author} {\bibfnamefont {K.}~\bibnamefont
  {Seki}}, \bibinfo {author} {\bibfnamefont {T.}~\bibnamefont {Shirakawa}},\
  and\ \bibinfo {author} {\bibfnamefont {S.}~\bibnamefont {Yunoki}},\
  }\bibfield  {title} {\enquote {\bibinfo {title} {Symmetry-adapted variational
  quantum eigensolver},}\ }\href {https://doi.org/10.1103/PhysRevA.101.052340}
  {\bibfield  {journal} {\bibinfo  {journal} {Phys. Rev. A}\ }\textbf {\bibinfo
  {volume} {101}},\ \bibinfo {pages} {052340} (\bibinfo {year}
  {2020})}\BibitemShut {NoStop}%
\bibitem [{\citenamefont {Mermin}(1990)}]{Mermin1990b}%
  \BibitemOpen
  \bibfield  {author} {\bibinfo {author} {\bibfnamefont {N.~D.}\ \bibnamefont
  {Mermin}},\ }\bibfield  {title} {\enquote {\bibinfo {title} {Extreme quantum
  entanglement in a superposition of macroscopically distinct states},}\ }\href
  {https://doi.org/10.1103/PhysRevLett.65.1838} {\bibfield  {journal} {\bibinfo
   {journal} {Phys. Rev. Lett.}\ }\textbf {\bibinfo {volume} {65}},\ \bibinfo
  {pages} {1838--1840} (\bibinfo {year} {1990})}\BibitemShut {NoStop}%
\bibitem [{\citenamefont {Collins}\ \emph {et~al.}(2002)\citenamefont
  {Collins}, \citenamefont {Gisin}, \citenamefont {Popescu}, \citenamefont
  {Roberts},\ and\ \citenamefont {Scarani}}]{Collins2002}%
  \BibitemOpen
  \bibfield  {author} {\bibinfo {author} {\bibfnamefont {D.}~\bibnamefont
  {Collins}}, \bibinfo {author} {\bibfnamefont {N.}~\bibnamefont {Gisin}},
  \bibinfo {author} {\bibfnamefont {S.}~\bibnamefont {Popescu}}, \bibinfo
  {author} {\bibfnamefont {D.}~\bibnamefont {Roberts}},\ and\ \bibinfo {author}
  {\bibfnamefont {V.}~\bibnamefont {Scarani}},\ }\bibfield  {title} {\enquote
  {\bibinfo {title} {{B}ell-type inequalities to detect true n-body
  nonseparability},}\ }\href {https://doi.org/10.1103/PhysRevLett.88.170405}
  {\bibfield  {journal} {\bibinfo  {journal} {Phys. Rev. Lett.}\ }\textbf
  {\bibinfo {volume} {88}},\ \bibinfo {pages} {170405} (\bibinfo {year}
  {2002})}\BibitemShut {NoStop}%
\bibitem [{\citenamefont {Alsina}\ \emph {et~al.}(2016)\citenamefont {Alsina},
  \citenamefont {Cervera}, \citenamefont {Goyeneche}, \citenamefont {Latorre},\
  and\ \citenamefont {\.{Z}yczkowski}}]{Alsina2016}%
  \BibitemOpen
  \bibfield  {author} {\bibinfo {author} {\bibfnamefont {D.}~\bibnamefont
  {Alsina}}, \bibinfo {author} {\bibfnamefont {A.}~\bibnamefont {Cervera}},
  \bibinfo {author} {\bibfnamefont {D.}~\bibnamefont {Goyeneche}}, \bibinfo
  {author} {\bibfnamefont {J.~I.}\ \bibnamefont {Latorre}},\ and\ \bibinfo
  {author} {\bibfnamefont {K.}~\bibnamefont {\.{Z}yczkowski}},\ }\bibfield
  {title} {\enquote {\bibinfo {title} {Operational approach to {B}ell
  inequalities: Application to qutrits},}\ }\href
  {https://doi.org/10.1103/PhysRevA.94.032102} {\bibfield  {journal} {\bibinfo
  {journal} {Phys. Rev. A}\ }\textbf {\bibinfo {volume} {94}},\ \bibinfo
  {pages} {032102} (\bibinfo {year} {2016})}\BibitemShut {NoStop}%
\bibitem [{\citenamefont {Garcia-Escartin}\ and\ \citenamefont
  {Chamorro-Posada}(2013)}]{Garcia-Escartin}%
  \BibitemOpen
  \bibfield  {author} {\bibinfo {author} {\bibfnamefont {J.~C.}\ \bibnamefont
  {Garcia-Escartin}}\ and\ \bibinfo {author} {\bibfnamefont {P.}~\bibnamefont
  {Chamorro-Posada}},\ }\bibfield  {title} {\enquote {\bibinfo {title} {swap
  test and hong-ou-mandel effect are equivalent},}\ }\href
  {https://doi.org/10.1103/PhysRevA.87.052330} {\bibfield  {journal} {\bibinfo
  {journal} {Phys. Rev. A}\ }\textbf {\bibinfo {volume} {87}},\ \bibinfo
  {pages} {052330} (\bibinfo {year} {2013})}\BibitemShut {NoStop}%
\bibitem [{Note5()}]{Note5}%
  \BibitemOpen
  \bibinfo {note} {Note that using the matrix form for $Y$ does \protect \emph
  {not} mean we have chosen the computation basis for the system
  qubits}\BibitemShut {NoStop}%
\bibitem [{\citenamefont {Alsina}\ and\ \citenamefont
  {Latorre}(2016)}]{Alsina2016a}%
  \BibitemOpen
  \bibfield  {author} {\bibinfo {author} {\bibfnamefont {D.}~\bibnamefont
  {Alsina}}\ and\ \bibinfo {author} {\bibfnamefont {J.~I.}\ \bibnamefont
  {Latorre}},\ }\bibfield  {title} {\enquote {\bibinfo {title} {Experimental
  test of mermin inequalities on a five-qubit quantum computer},}\ }\href
  {https://doi.org/10.1103/PhysRevA.94.012314} {\bibfield  {journal} {\bibinfo
  {journal} {Phys. Rev. A}\ }\textbf {\bibinfo {volume} {94}},\ \bibinfo
  {pages} {012314} (\bibinfo {year} {2016})}\BibitemShut {NoStop}%
\bibitem [{\citenamefont {Gu}, \citenamefont {Somma},\ and\ \citenamefont
  {{\c{S}}ahino{\u{g}}lu}(2021)}]{Gu2021}%
  \BibitemOpen
  \bibfield  {author} {\bibinfo {author} {\bibfnamefont {S.}~\bibnamefont
  {Gu}}, \bibinfo {author} {\bibfnamefont {R.~D.}\ \bibnamefont {Somma}},\ and\
  \bibinfo {author} {\bibfnamefont {B.}~\bibnamefont {{\c{S}}ahino{\u{g}}lu}},\
  }\bibfield  {title} {\enquote {\bibinfo {title} {Fast-forwarding quantum
  evolution},}\ }\href {https://doi.org/10.22331/q-2021-11-15-577} {\bibfield
  {journal} {\bibinfo  {journal} {Quantum}\ }\textbf {\bibinfo {volume} {5}},\
  \bibinfo {pages} {577} (\bibinfo {year} {2021})}\BibitemShut {NoStop}%
\bibitem [{\citenamefont {Yoshida}(1990)}]{Yoshida1990}%
  \BibitemOpen
  \bibfield  {author} {\bibinfo {author} {\bibfnamefont {H.}~\bibnamefont
  {Yoshida}},\ }\bibfield  {title} {\enquote {\bibinfo {title} {Construction of
  higher order symplectic integrators},}\ }\href
  {https://doi.org/10.1016/0375-9601(90)90092-3} {\bibfield  {journal}
  {\bibinfo  {journal} {Phys. Lett. A}\ }\textbf {\bibinfo {volume} {150}},\
  \bibinfo {pages} {262--268} (\bibinfo {year} {1990})}\BibitemShut {NoStop}%
\end{thebibliography}%

\end{document}